\documentclass{article}
\usepackage[utf8]{inputenc}
\usepackage[margin=1in]{geometry}
\usepackage{subcaption}
\usepackage{graphicx}
\usepackage{amsmath}
\usepackage[hyphens]{url}
\usepackage{hyperref}
\hypersetup{colorlinks=false,breaklinks=true,hidelinks}
\usepackage{cleveref}
\usepackage{apacite}
\usepackage{natbib}
\usepackage{authblk}
\usepackage{multirow}
\usepackage{booktabs}
\usepackage{makecell}
\graphicspath{{figures/}}
\usepackage{setspace}
\title{Long-term marine acoustic and seismic monitoring using distributed acoustic sensing and deep learning}

\author[1,2$\ast$]{Chun Zhang}

\author[1,2]{Weiqiang Zhu}
\author[1,2]{Barbara A. Romanowicz}
\author[1,2]{Richard M Allen}
\author[3]{Kenichi Soga}
\author[4]{Yuxin Wu}

\affil[1]{Department of Earth \& Planetary Science, University of California, Berkeley}
\affil[2]{Berkeley Seismology Laboratory, University of California, Berkeley}
\affil[3]{Department of Civil \& Environmental Engineering, University of California, Berkeley}
\affil[4]{Earth and Environmental Sciences Area, Lawrence Berkeley National Laboratory}
\affil[*]{Corresponding author. Email: chun.zhang@berkeley.edu}

\date{}

\begin{document}

\maketitle

\begin{abstract}

The ocean remains one of the least instrumented parts of Earth, and many geophysical, biological, and anthropogenic signals go undetected for lack of instrumentation. Distributed acoustic sensing (DAS) can transform submarine fiber-optic cables into dense seafloor sensor arrays, but extracting diverse signals from massive DAS recordings remains challenging. Here we present DASNet, a deep learning framework that detects, classifies, and picks arrival times of diverse marine signals in continuous DAS data. Applied to nearly four years of Seafloor Fiber-Optic Array in Monterey Bay recordings, DASNet identifies more than 620,000 events. These detections reveal local earthquakes; distant earthquake- and volcanic-eruption-generated T-waves from the southwestern Pacific and mid-ocean ridge systems; more than 510,000 blue and fin whale calls with seasonal and interannual variability consistent with hydrophone records; and vessel traffic near the cable. Together, these results show that submarine fiber-optic cables combined with deep learning enable scalable, high-resolution ocean monitoring.
\end{abstract}

\section{Introduction}
The open ocean carries acoustic and seismic signals from tectonic, biological, and anthropogenic sources, spanning timescales from the seconds of a seismic arrival to the seasonal cycles of marine life.
Natural hazards such as earthquakes, submarine landslides, and volcanic eruptions reshape the seafloor and can generate tsunamis that reach distant coastlines \citep[]{carvajal2022worldwide, harbitz2006mechanisms, kanamori1972mechanism, omira2022global}, with cascading effects on benthic habitats and industrial infrastructure. 
Whales and other megafauna communicate using low-frequency sound, while expanding shipping traffic, offshore wind farms, and seabed infrastructure impose persistent acoustic noise in the same waters \citep[]{roman2014whales, ryan2022oceanic, hildebrand2009anthropogenic, ryan2025audible}. 
Despite the scientific and societal importance of these processes, routine offshore observations remain sparse and episodic. Conventional ocean-bottom seismometers and hydrophones are costly to deploy, limited in spatial and temporal coverage, and typically restricted to short-term campaigns \citep{toomey2014cascadia, suetsugu2014broadband}, leaving hazards under-detected, biodiversity trends under-sampled, and anthropogenic impacts under-quantified.

Distributed acoustic sensing (DAS) offers a promising direction to address this observational gap by converting existing submarine fiber-optic cables into dense sensor arrays. DAS measures minute strain or strain-rate along the fiber based on Rayleigh backscattering, enabling continuous, high-resolution monitoring at reduced deployment costs \citep{zhan2020distributed, lindsey2019illuminating, lindsey2021fiber, williams2019distributed, romanowicz2023seafoam}. 
Over the past several years, DAS has been used to detect earthquakes, marine mammal vocalizations, and anthropogenic signals such as vessel traffic and infrastructure operations, demonstrating its potential for integrated, long-duration ocean observation \citep[]{williams2019distributed, sladen2019distributed, lior2021detection, williams2022surface, landro2022sensing, bouffaut2022eavesdropping, shen2024ocean, xiao2024detection, gou2025leveraging}. 
However, realizing the full potential of DAS poses substantial challenges. Data quality varies with cable design, installation conditions, and coupling to the surrounding medium \citep{wang2018ground, lindsey2020broadband, muir2022wavefield}, so DAS recordings are frequently dominated by noise or exhibit low sensitivity to certain sources. In addition, DAS recordings often contain a superposition of diverse wavefields. Seismic phases, ocean acoustic signals, and anthropogenic noise occupy frequencies from millihertz to hundreds of hertz and propagate at different apparent velocities and coherence lengths, complicating their isolation and attribution. At the same time, a single DAS array can generate terabytes of data per day from thousands of channels, making manual inspection impractical for event detection and characterization \citep{ma2023machine, romanowicz2023seafoam}. These challenges motivate the development of automated and scalable analysis methods for DAS.

Recent advances in machine learning provide powerful tools for automated and scalable DAS data analysis. Neural networks can learn representations directly from raw waveforms, enabling efficient and accurate detection across diverse signal types. 
For example, machine learning has reshaped earthquake-monitoring tasks, such as seismic phase picking \citep[]{zhu2019phasenet, mousavi2020earthquake}, phase association \citep[]{ross2019phaselink, zhang2019rapid, zhu2022earthquake}, and waveform denoising \citep[]{zhu2019seismic}. However, these advances have largely relied on decades of high-quality labeled data from traditional seismic networks. No comparable training corpus exists for DAS despite its rapid expansion. 
Moreover, DAS recordings exhibit characteristics that differ from conventional seismograms, including strong directional sensitivity, variable cable coupling, coherent noise patterns, frequent superposition of multiple sources, and non-standard data formats \citep{wang2018ground, lindsey2019illuminating, muir2022wavefield}. These factors require specialized approaches, not simply adaptations of existing seismic models. 
Supervised DAS models trained on carefully annotated subsets have demonstrated the ability to detect microseismic and other weak signals that conventional methods often miss, but they remain constrained by labeling cost and limited generalization across arrays \citep{huot2022detection, ma2023machine, yu2024daseventnet, 10.1093/gji/ggag061}. 
Unsupervised and self-supervised approaches instead learn stable representations from continuous DAS streams without manual labels, though they still require a downstream task and a small labeled set to transfer learned features into usable detections \citep{zhu2023seismic, van2021self, ding2025dasformer, shen2026unsupervised}. 
Despite these advances, few existing methods generalize across multiple DAS signal types simultaneously due to both the complexity of DAS wavefields and the specialized training that each signal class demands.
This represents a critical gap for submarine DAS, where a single cable records seismic, bioacoustic, and anthropogenic signals. A unified detection framework would enable integrated analysis across geophysics, oceanography, and environmental monitoring. 

Here we present DASNet, a deep learning framework that automatically detects, classifies, and picks the arrival times of diverse marine signals in DAS data. 
For each event, DASNet produces a bounding box that marks its extent in time and across channels, a class label, and a segmentation mask from which the arrival-time curve is read.
The model is trained using a semi-supervised pipeline that iteratively improves on continuous recordings, making it practical for long DAS deployments where labeled data are limited but unlabeled data are abundant. We apply DASNet to nearly four years of SeaFOAM data from Monterey Bay \citep[]{romanowicz2023seafoam} and show that a unified model can support four representative marine monitoring tasks: (i) detecting local earthquakes and measuring phase arrival times for source localization; (ii) identifying T-waves generated by distant earthquakes and estimating source back-azimuths via beamforming; (iii) cataloging baleen whale calls to reveal seasonal and interannual behavioral patterns; and (iv) tracking anthropogenic signals, such as vessel traffic, near the cable. 
These examples demonstrate that DASNet can transform continuous fiber-optic recordings into structured multi-class event catalogs without task-specific preprocessing. 
As submarine fiber-optic networks continue to expand worldwide, coupling DAS with scalable machine learning could expand our capacity to observe seismicity, marine ecosystems, and human activity across ocean basins that remain beyond the reach of traditional instrumentation.

\section{Methods}
\begin{figure}[htbp]
    \centering
    \begin{subfigure}{1\textwidth}
        \centering
        \includegraphics[width=\textwidth]{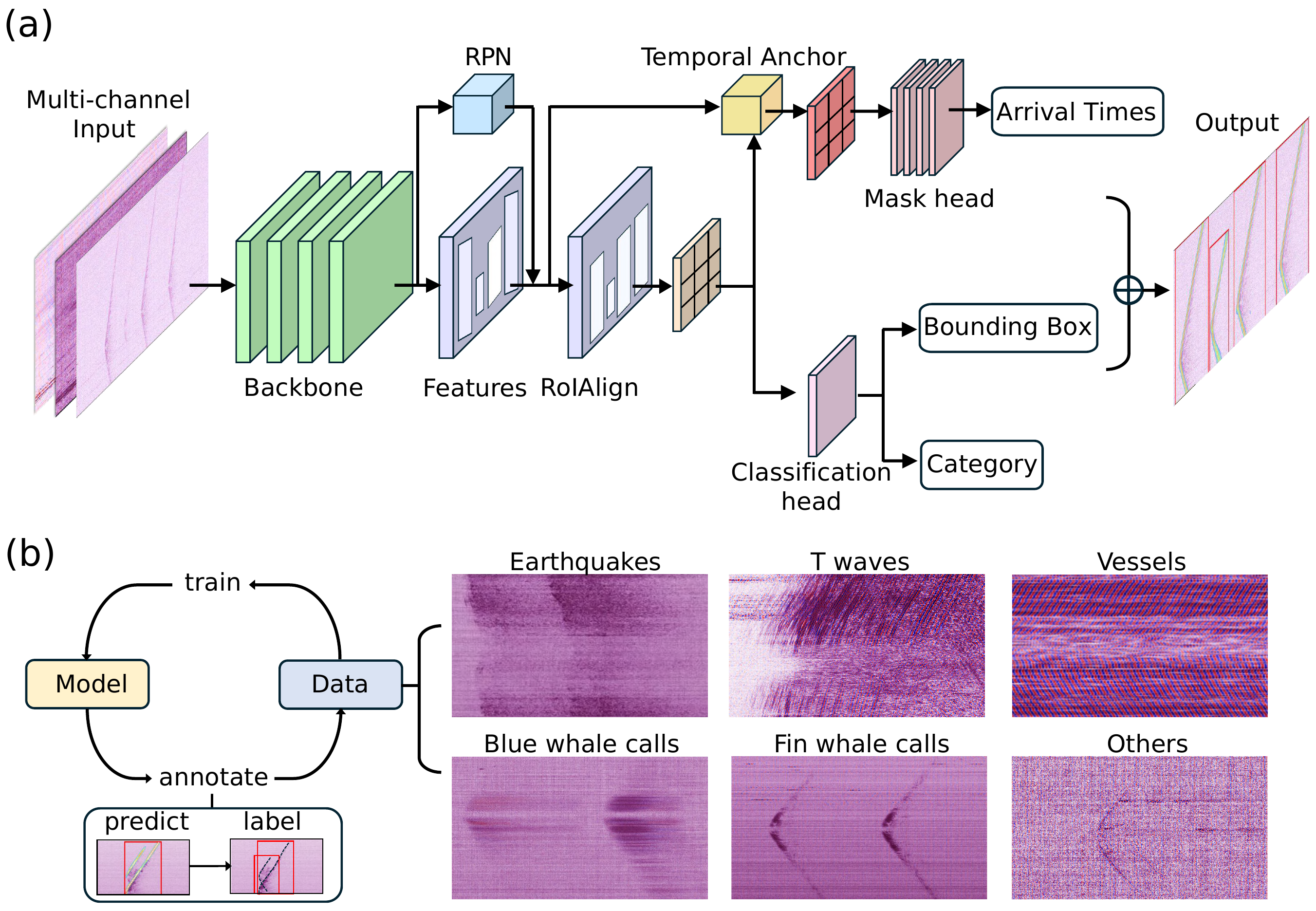}
        \label{fig:subfig1}
    \end{subfigure}
    \caption{DASNet framework overview. (a) Model architecture with the feature extraction backbone, region proposal network, and output branches for classification, bounding box regression, and arrival-time mask prediction. (b) Semi-supervised data engine workflow for iterative dataset expansion and model refinement. Only a subset of whale call categories is shown here; the complete set is presented in \Cref{fig:example_and_eval}.}
    \label{fig:model_engine}
\end{figure}

\subsection{Neural network architecture}

DASNet (\Cref{fig:model_engine}a) adapts Mask R-CNN, a deep learning framework designed for simultaneous object detection and instance segmentation \citep{he2017mask}, to detect and measure diverse seismic and acoustic signals in DAS recordings. For each detected instance, the model produces three outputs: (1) a bounding box delineating the spatial-temporal extent of the signal, (2) a classification label with an associated confidence score, and (3) a probability mask over channels and time that highlights arrival energy and supports arrival-time estimation. By treating each event as a discrete object in the 2D channel-time domain, this approach departs from conventional channel-wise detectors and naturally handles monitoring conditions where multiple sources overlap in time and across channels. 

The DASNet architecture consists of five primary components: a feature extraction backbone, a Region Proposal Network (RPN), a Region of Interest Align (ROIAlign) module, a Temporal Anchor Network, and multiple output branches for classification, bounding box regression, and mask generation. 
It follows the two-stage design of Mask R-CNN that yields accurate instance masks and stable localization at moderate computational cost. 
In the first stage, a convolutional backbone with a Feature Pyramid Network (FPN) extracts multiscale representations that capture both local waveform characteristics and array-wide patterns. We employ ResNet-50 \citep{he2016deep} with FPN, which provides a good balance between accuracy and training stability for long sequences. Alternative backbones such as Vision Transformer, Swin Transformer or ConvNeXt \citep{dosovitskiy2020image, liu2021swin, liu2022convnet} may improve accuracy but were not explored here, as the present effort emphasizes establishing a reproducible baseline. 
Then, region proposals are generated by the RPN and are refined by ROIAlign to preserve alignment between features and the underlying channel-time grid. 
In the second stage, three prediction branches operate on each proposed region. The classification branch assigns an event type, the regression branch refines the spatiotemporal bounding box, and the multi-stream mask branch predicts a high-resolution probability mask from which arrival-time trajectories are extracted.
Because DAS signals can span much longer temporal durations than spatial extent, they differ substantially from conventional image-based tasks. To address this spatiotemporal aspect-ratio imbalance, we add a Temporal Anchor Network to the mask prediction branch. For each detection proposal, it predicts a temporal anchor from the shared RoI features, adjusts the mask RoI along the time axis, and re-pools features before the multi-stream mask heads, yielding a more appropriate aspect ratio for mask learning. The corresponding formulation is given in Text S1 of the Supplementary Information.

The model architecture has been optimized to balance computational efficiency and performance, given the high-resolution nature of DAS data. We added additional downsampling in the backbone by modifying the stride and downsampling behavior of early convolutional blocks. This reduces the resolution of intermediate feature maps and substantially lowers computational cost while preserving event-scale semantic information required for detection and segmentation. The FPN was configured with a reduced channel dimension compared to the standard Mask R-CNN implementation, further improving memory efficiency. 
Detailed model configurations and implementation details are provided in Text~S1 of the Supplementary Information. 
Beyond architectural modifications, the input representation was designed to enhance feature extraction across the diverse spectral characteristics of DAS signals. We employ a multi-channel input strategy that encodes information from three frequency bands: (1) raw strain-rate, which retains the full broadband signal to preserve critical features; (2) bandpass-filtered strain-rate (2-10 Hz), which highlights low-frequency signals including local earthquakes and T-waves; (3) high-pass filtered strain-rate ($>$10 Hz), which emphasizes high-frequency signals, such as whale calls and vessel noise. 
This structured input design enables the model to effectively learn the spectral characteristics of different signal types and enhances its ability to distinguish among geophysical, biological, and anthropogenic sources.

\subsection{Semi-supervised data engine}

The rapid expansion of global DAS networks has generated vast data volumes \citep[]{wuestefeld2024global}. However, the lack of a comprehensive annotated dataset poses a major challenge for supervised learning approaches. As fully manual annotation is impractical given the scale of DAS data, we apply a semi-supervised data engine to facilitate annotation and dataset expansion (\Cref{fig:model_engine}b). The data engine operates in two distinct stages: (1) an initial manual annotation stage, and (2) a semi-automatic stage where model-generated predictions assist human annotation.

In the first stage, a subset of DAS data is manually reviewed and labeled to establish a preliminary dataset. Each annotated instance includes three components: a bounding box capturing the full signal extent, a categorical label indicating the signal type, and a manually picked signal arrival-time curve. The mask representation is generated as a Gaussian distribution centered on the picked time curve. This probabilistic representation mitigates inconsistencies introduced by human labeling errors \citep{zhu2019phasenet} and provides the model with a structured learning target. 
We split this dataset into training and test sets at an 8:2 ratio. Each input segment has an input size of 12,000$\times$2,845 (60~s at 200~Hz sampling rate across 2,845~channels) and is filtered to produce multi-channel model input, with each segment normalized independently. To expand data diversity, we applied stochastic data augmentation: superposing synthetic noise to improve robustness to low SNR events, applying random spatial flips and slight stretching (temporal and spatial resampling) to enhance generalization across different DAS configurations, and stacking multiple signals to sharpen the model’s ability to separate overlapping instances. 
The model trained on this augmented dataset then served as the starting point for the second phase. For training, we used the AdamW optimizer with a weight decay of 1$\times$10$^{-4}$. The initial learning rate was set to 1$\times$10$^{-4}$ with a cosine decay schedule and a linear warm-up over the first epoch. The models were trained for 20 epochs using an effective batch size of 8.

Once the initial model was trained, the data engine transitioned into the semi-supervised iterative phase, a label-refinement strategy that has also enabled earthquake detection in sparse ocean-bottom seismic data \citep{xi2024deep}.
The model was deployed on continuous DAS recordings, generating predictions despite limited initial accuracy. The predicted bounding boxes and categorical labels were retained, and the signal arrival-time curves were extracted from the probability distribution masks. These preliminary annotations were then manually reviewed and refined before being incorporated into the dataset. The updated dataset was subsequently used to retrain the model, enhancing its detection and measurement capabilities. Throughout this process, we kept the dataset splitting ratio as 8:2 and applied the same training strategy. 
This iterative process progressively improved the model performance, which in turn accelerated subsequent dataset expansion in a self-reinforcing cycle. For example, the annotation efficiency increases with each iteration. Early stages require significant manual correction of low-quality model predictions, making annotation time-consuming, whereas later stages produce automatic labels of sufficient quality, requiring only minimal human intervention. 
Eventually, the data engine achieves a near-automatic state, enabling large-scale dataset expansion with minimal manual effort.

\subsection{The SeaFOAM dataset}

\begin{figure}[htbp]
    \centering
    \begin{subfigure}{0.71\textwidth}
        \centering
        \includegraphics[width=\textwidth]{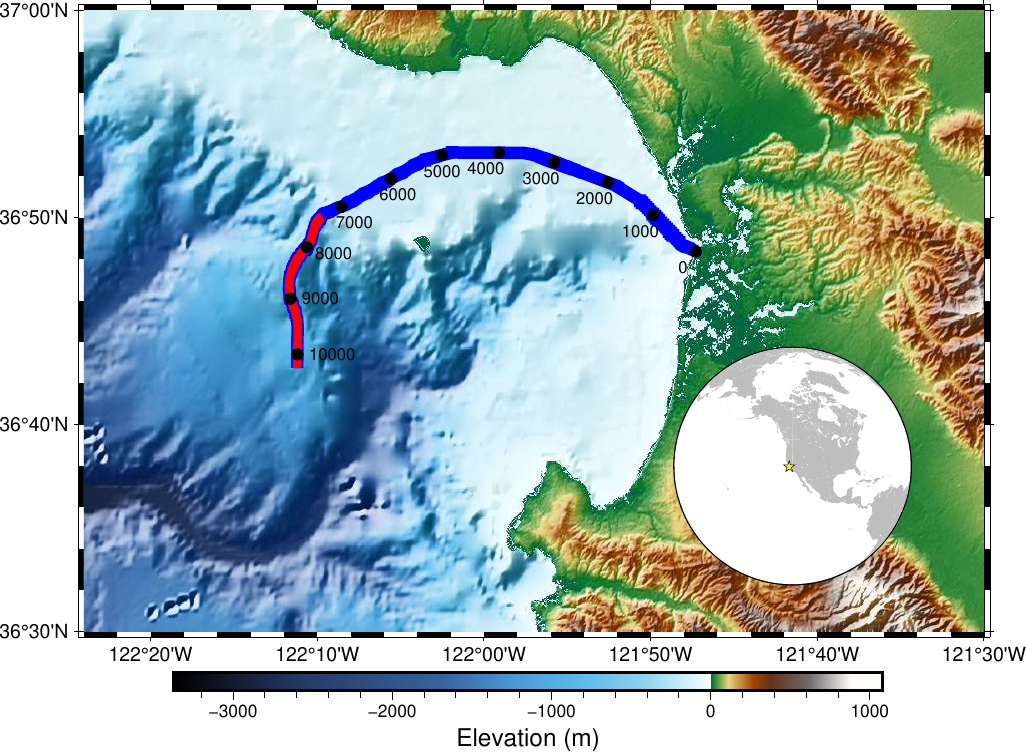}
        \label{fig:subfig1}
    \end{subfigure}
    \hfill %
    \begin{subfigure}{0.25\textwidth}
        \centering
        \includegraphics[width=\textwidth]{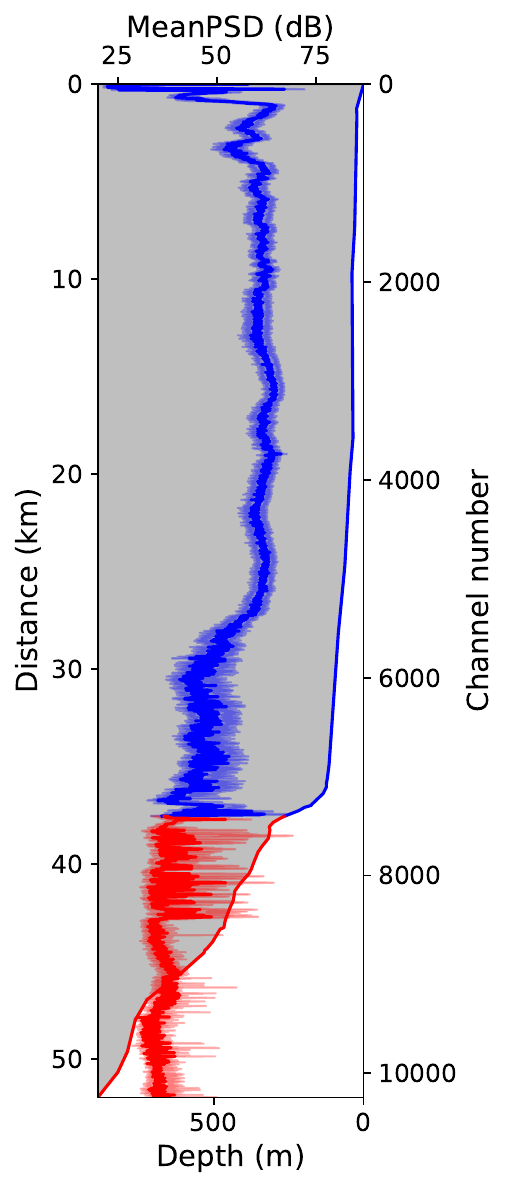}
        \label{fig:subfig2}
    \end{subfigure}

    \caption{SeaFOAM DAS array configuration and data characteristics. (a) Location of the SeaFOAM DAS array. The red line denotes the section used in this study, whereas the blue line indicates the unused portion. Black dots mark every 1,000 channels along the cable from the interrogator unit. The inset in the lower right shows the location of the cable in Monterey Bay, California. (b) Depth profile of the underwater DAS channels, overlaid with the daily median power spectral density (PSD) for each channel on 25 November 2023 (solid line). Shaded areas represent the 10th - 90th percentiles of PSDs, illustrating channel-to-channel signal variation.}
    \label{fig:SeaFOAM}
\end{figure}

In this study, we conducted the entire model iteration, signal detection, and arrival-time picking workflow on the SeaFOAM (Seafloor Fiber-Optic Array in Monterey Bay) dataset \citep{romanowicz2023seafoam}. The SeaFOAM experiment deploys DAS on a 52~km submarine cable offshore Monterey Bay, California, with data acquisition beginning on July 21, 2022. The system records data at a 200~Hz sampling rate from 10,245 channels with 5.1~m channel spacing. The SeaFOAM DAS data capture a wide range of seismic and oceanic signals, including local and teleseismic events, ocean currents, water waves, ambient noise, and marine mammal vocalizations. The water depth varies significantly along the cable, so the background noise characteristics change accordingly, with much higher noise levels in shallow water. For this reason, we focused on the deep-water section (channels 7,400-10,245, $>$200~m depth and $\sim$37.6-52~km distance from shore) with better signal quality (\Cref{fig:SeaFOAM}).

We initiated the first phase of the data engine using records from August 2022, because this period coincides with known whale activity offshore of California \citep[]{wiggins2005blue}. We conducted a rough manual search for signals of interest, including earthquake signals and several types of whale calls. This small dataset was used to train the first model, which then served as the starting point for the second phase of iteration. In subsequent rounds, we applied the model to continuous data to expand the dataset and retrain updated models. Early iterations applied the model to shorter time spans for rapid feedback, and later iterations extended to longer archives as detection quality stabilized. 
At each stage, we performed time-uniform sampling to select a representative and diverse subset of predictions for manual review and model refinement. 
Given the substantial class imbalance among signal types, we controlled the number of samples selected per class to ensure adequate representation and applied oversampling to amplify underrepresented classes. Through successive iterations, the model's performance improved significantly, ultimately yielding more than 620,000 detected events across the full dataset. 

\section{Results}
\subsection{Multi-class signal detection and arrival-time measurement}

\begin{figure}[htbp]
    \centering
    \includegraphics[width=0.8\textwidth]{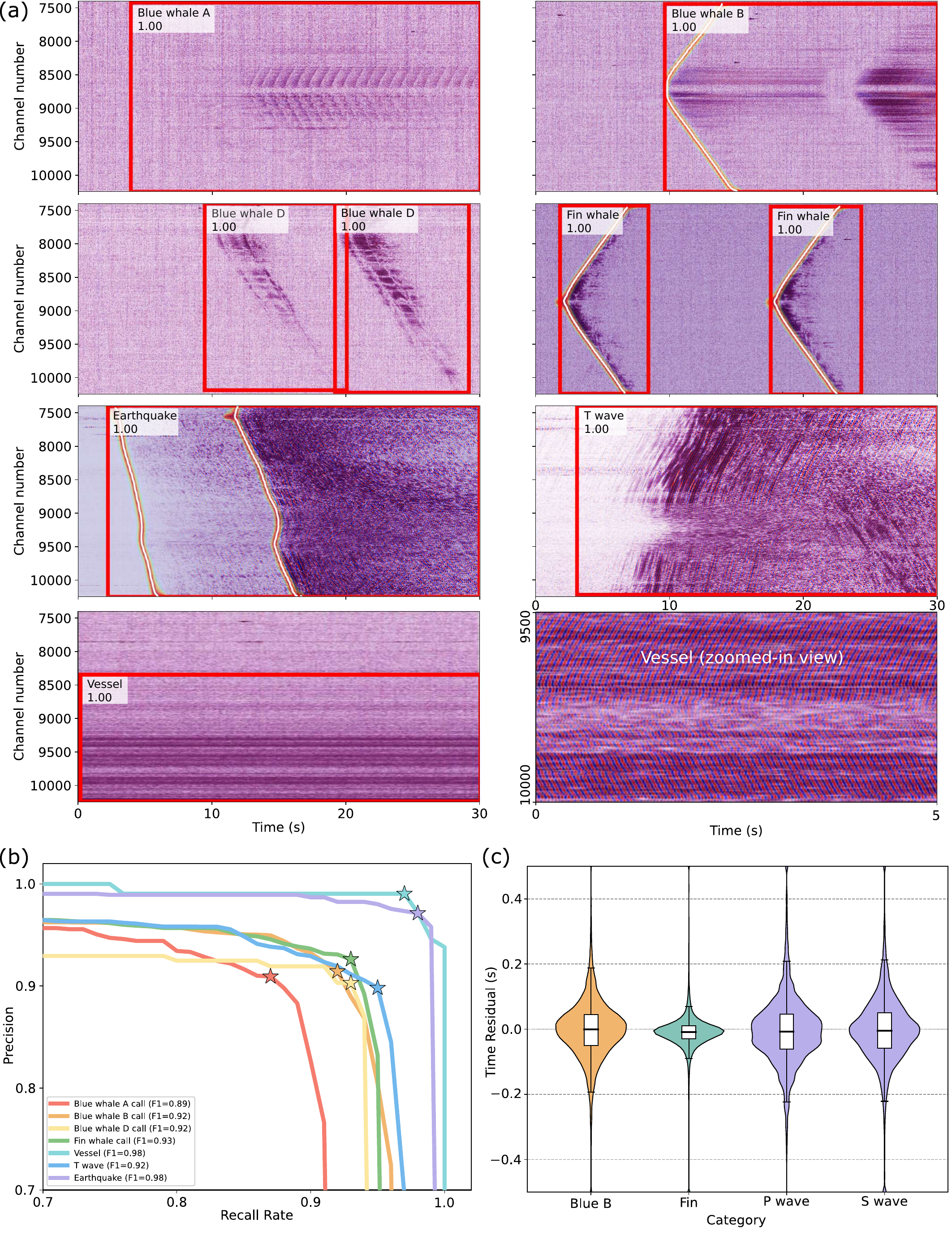}
    \caption{Model performance evaluation and signal examples. (a) Representative examples of each signal type detected in the SeaFOAM dataset. Red boxes mark detected events, the overlaid mask traces the picked arrival time, and the label gives the predicted class and its confidence. Earthquake and T-wave segments are bandpass filtered between 2–10 Hz, while the remaining signals are highpass filtered above 10 Hz to enhance relevant features. The corresponding spectrograms are shown in Figure S1. (b) Precision-recall (PR) curves for all signal categories evaluated on the test set at an intersection-over-union (IoU) threshold of 0.5. Stars on each curve mark the point corresponding to the highest F1 score. (c) Distribution of arrival-time residuals on the test set, computed as the difference between predicted and manually labeled arrival times. Predicted arrivals are extracted as the peak of the model’s probability mask using a threshold of 0.5. Boxplots display the median (bold horizontal line), interquartile range (IQR; box boundaries), and whiskers extending to 1.5 $\times$ IQR from the quartiles. Outliers beyond this range are omitted for clarity.}
    \label{fig:example_and_eval}
\end{figure}

Automated multi-class signal detection in marine DAS data poses significant challenges because the signals of interest span a wide range of spatiotemporal scales and spectral characteristics (Figure S1). For example, local earthquake signals primarily exhibit energy below 10~Hz, while teleseismic arrivals are typically strongest below 1~Hz. Marine mammal vocalizations occupy distinct frequency bands. Fin whale calls appear around 15-25~Hz, while blue whale calls contain multiple harmonics from 10~Hz upward. Vessel-generated noise is often characterized by narrowband peaks above 10~Hz. 
This broad spectral diversity limits both traditional signal processing methods and single-task machine learning models, which struggle to generalize across signal categories.
By taking the different frequency bands as separate input channels, DASNet can effectively process complex DAS data with diverse spectral features. \Cref{fig:example_and_eval}(a) shows representative examples of these signal types observed along the DAS array. Although acoustic and seismic signals vary widely in frequency content, amplitude, and temporal structure, DASNet successfully classifies, localizes (via bounding boxes), and measures arrival times for these events. The applicable categories include seismic events (P-, S-, and T-waves), marine mammal calls (blue whale A, B, and D calls; fin whale calls), and anthropogenic sources (vessel noise).

We evaluated DASNet on a held-out test set constructed from SeaFOAM recordings. The model reliably detected signals across all categories, with best F1 scores ranging from 0.89 to 0.98 for most categories
(\Cref{fig:example_and_eval}b). 
This performance is largely attributable to the rich and diverse training dataset constructed through semi-supervised iterations (see Methods), which enabled the model to capture representative characteristics of each signal type. Manual inspection of failed predictions revealed that most missed detections and false positives arose from inherently weak signals, strong background noise, or truncated waveforms at segment boundaries. 
For example, blue whale A call detection (best F1 = 0.89) was affected by faint signals from distant sources, whereas earthquake detections (best F1 = 0.98) benefited from the strong and sustained character of phase arrivals. 
The reported F1 scores correspond to the optimal operating point on each precision–recall curve. More broadly, the detection-score threshold controls a trade-off between detection quantity and quality. For subsequent analyses, we adopted a relatively stringent threshold (0.8) to retain only high-quality detections. However, this conservative choice does not imply that lower-score detections are spurious. On continuous data, low-score detections co-vary temporally with high-confidence ones rather than emerging during quiet periods (Figure S2), indicating that they predominantly correspond to genuine but weaker or noisier signals. 

In addition to detecting acoustic and seismic events, precise measurement of their arrival times 
is essential for scientific analyses such as source localization and subsurface structure imaging. 
We evaluated the arrival-time accuracy using manually annotated or reviewed arrival picks for signals with clear and consistent onsets, such as blue whale B calls, fin whale calls, and earthquake P- and S-waves. 
\Cref{fig:example_and_eval}(c) shows that DASNet achieved high timing precision for marine mammal vocalizations, with mean absolute errors of 0.07~s (standard deviation 0.10~s) for blue whale B calls, and a mean of 0.04~s (standard deviation 0.08~s) for fin whale calls. 
Seismic phases exhibited lower timing precision, with a mean of approximately 0.08~s (standard deviation 0.11~s) for P-waves, and a mean of 0.08~s (standard deviation of 0.12~s) for S-waves. 
Overall, the arrival-time precision is sufficient for downstream analyses, such as event localization and subsurface structure imaging, across different acoustic and seismic signal types.

The experiments above demonstrate that DASNet can effectively detect and measure marine acoustic and seismic signals, expanding DAS's utility across diverse scientific disciplines. In the subsequent sections, we apply the trained model to nearly four years of continuous SeaFOAM recordings to explore its potential for several environmental and geophysical monitoring applications.

\subsection{Local earthquake detection and localization} 
Seismic monitoring in offshore regions has long been limited by sparse instrumentation coverage. 
DAS presents a promising approach to address this data gap by providing dense spatial sampling along existing submarine cables \citep[]{williams2019distributed, fernandez2022seismic}. 
Realizing this potential requires automated methods for phase detection and arrival-time picking that can operate reliably on continuous, high-volume DAS recordings.
The earthquake detection and phase-picking capabilities of DASNet could therefore enable DAS to improve the coverage and resolution of offshore earthquake monitoring and subsurface imaging \citep{gou2025leveraging, lior2022imaging}.
\begin{figure}[htbp]
    \centering
    \includegraphics[width=0.95\textwidth]{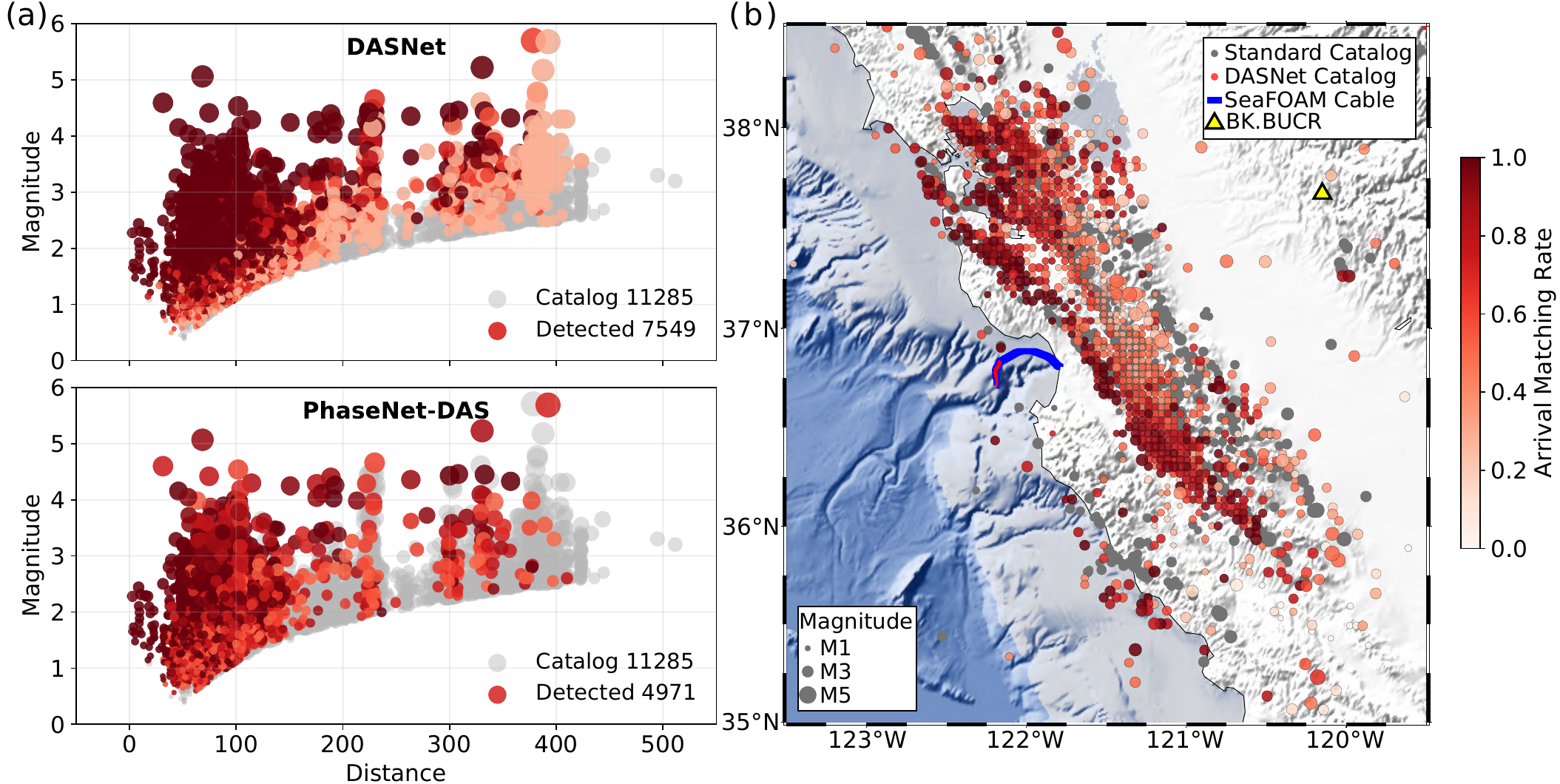}
    \label{fig:compare}
    \caption{Earthquake detection and localization results. (a) Comparison of the earthquake detection performance between DASNet and PhaseNet-DAS as a function of event magnitude and epicentral distance, where the distance is computed relative to the average location of the DAS cable. Color denotes the arrival matching rate, the fraction of DAS channels with picks consistent with theoretical arrivals, and symbol size scales with earthquake magnitude. Gray points are all cataloged events. (b) Earthquake epicenters determined by combining P- and S-phase arrival times picked by DASNet on SeaFOAM DAS and by PhaseNet on the conventional seismic station BK.BUCR.}
    \label{fig:earthquake_monitor}
\end{figure}
    
To evaluate DASNet’s performance in detecting real earthquake signals, 
we selected earthquakes from the Northern California Seismic Network (NCSN) catalog that occurred between 2022 and 2026 within a region around the cable, bounded by coordinates 124$^\circ$00$^\prime$~W, 117$^\circ$46$^\prime$~W and 34$^\circ$30$^\prime$~N, 40$^\circ$00$^\prime$~N. 
Due to the limited spatial coverage of the DAS cable compared to the regional seismic network, signals from distant small-magnitude earthquakes are expected to be too weak to be detected by DAS. 
Thus, we filtered the catalog based on an approximate magnitude-distance scaling relation \citep{yin2023earthquake}.
To establish a reference baseline, we applied the pre-trained PhaseNet-DAS model to the same time period, using a low confidence threshold (0.3) to maximize recall at the expense of increased false positives. 
For DASNet, an event was considered detected when the theoretical arrival window was enclosed by a predicted event bounding box, independent of whether all individual phase arrivals were successfully picked.
As shown in \Cref{fig:earthquake_monitor}(a), DASNet detects more of the cataloged events than PhaseNet-DAS (7,549 versus 4,971 of 11,285), particularly for small-magnitude events near the DAS array. The arrival matching rate, defined as the fraction of channels whose picks agree with the theoretical arrivals, exhibits a systematic magnitude–distance dependence, with larger and closer earthquakes generally showing higher agreement. This enhanced performance partly arises from differences in model design. PhaseNet-DAS directly detects phase arrivals from continuous data, while DASNet first detects an event via a bounding box, then picks precise phase arrivals within the detected window. This two-stage design lets DASNet exploit coherent signal features across longer temporal windows and recover faint earthquake signals whose individual phase arrivals are difficult to identify.

Using these DASNet-derived picks, we further estimated earthquake source locations. The resulting epicenters broadly reproduce the spatial patterns of cataloged seismicity (Figure S3), validating the reliability of DASNet picks. However, some event locations remain poorly constrained, particularly for events at greater distances from the cable, due to limited azimuthal coverage inherent to a single DAS array and the reduced availability of reliable phase picks.
Combining DAS with conventional seismic stations can directly address this limitation. 
As an example, we integrated the DASNet picks with PhaseNet picks from a broadband seismic station
(BK.BUCR). Compared with the DAS-only locations (Figure S3), the joint inversion yields more stable location estimates and improved agreement with the cataloged seismicity distribution (\Cref{fig:earthquake_monitor}b). The arrival matching rate also exhibits clear spatial variations, with generally higher values observed to the north and southeast of the cable. This pattern likely reflects the directional sensitivity of the cable geometry. These results suggest 
that future efforts combining DAS with seismic networks, leveraging multiple cables, or deploying cables with optimized geometries could improve both localization accuracy and spatial coverage for offshore earthquake monitoring.

\subsection{T-wave monitoring and source attribution}

T-waves are acoustic phases generated when seismic body waves convert into acoustic energy at the seafloor and propagate efficiently through the ocean’s Sound Fixing and Ranging (SOFAR) channel over thousands of kilometers \citep{tolstoy1950t}. 
Because they are commonly excited by distant offshore earthquakes or volcanic eruptions, T-waves offer a valuable window to monitor offshore and deep-ocean seismicity in regions poorly covered by conventional seismic networks \citep{fox2001monitoring, smith2002hydroacoustic, tepp2021seismo, wech2018using}. Furthermore, T-waves from stable and repeatable source-receiver paths also enable ocean thermometry through time-lapse measurements of sound-speed variations \citep{wu2020seismic}. However, seismic stations capable of capturing T-waves remain sparsely distributed across the ocean. DAS has proven effective for T-wave detection \citep{shen2024ocean}. In addition to enabling long-term continuous monitoring, the dense spatial sampling of DAS arrays allows source azimuth to be constrained via beamforming techniques \citep[]{van2021evaluating}.
\begin{figure}[htbp]
    \centering
    \includegraphics[width=1\textwidth]{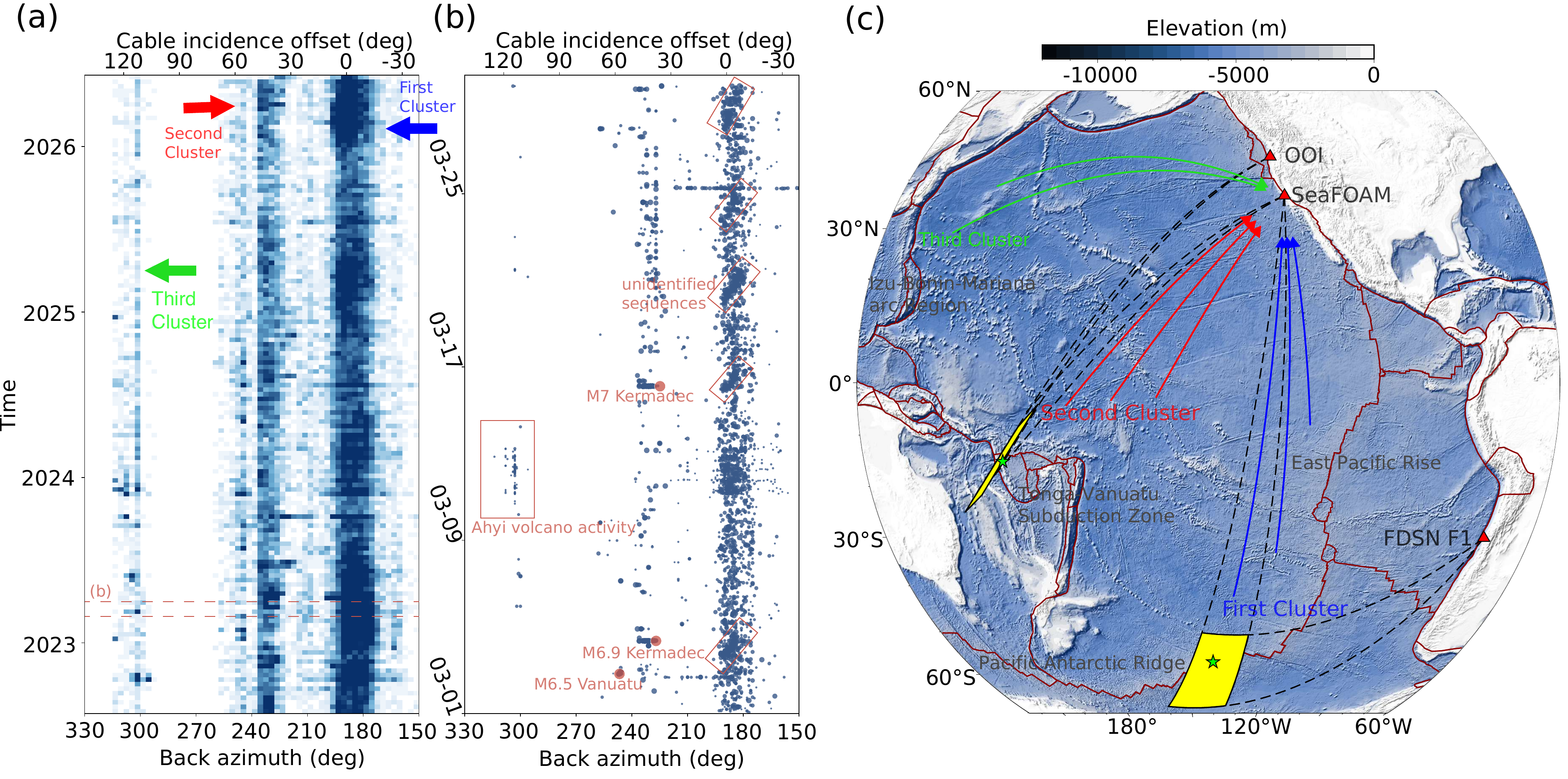}
    \caption{T-wave catalog and multi-array source localization.
    (a) Four-year T-wave catalog containing over 83,000 detections, shown as a function of back azimuth and time. Color indicates the relative density of detections. Three persistent azimuthal clusters are observed throughout the observation period, corresponding to the source regions summarized in (c). The upper axis shows the incidence offset between the inferred arrival direction and the mean cable orientation. The dashed box marks the time interval shown in (b). (b) Expanded view of a selected time interval illustrating individual T-wave detections. Several sequences coincide with major submarine earthquakes 
    (Mw 6.5, 2023-03-02 18:04:30 UTC, 15.377$^{\circ}$S 166.391$^{\circ}$E; Mw 6.9, 2023-03-04 06:41:23 UTC, 29.502$^{\circ}$S 178.797$^{\circ}$W; Mw 7.0, 2023-03-16 00:56:00 UTC, 30.174$^{\circ}$S 176.205$^{\circ}$W), aftershock activity, and independently reported volcanic unrest episodes \citep{GVP_Ahyi_2023}. The tilted boxes highlight recurring unidentified T-wave sequences that exhibit systematic azimuthal migration through time. Point size scales with the corresponding beam power. (c) Map showing the locations of the SeaFOAM, OOI, and RESIF F1 DAS installations \citep[]{romanowicz2023seafoam, shi2024multiplexed, https://doi.org/10.15778/resif.f1}. Arrows indicate the dominant azimuths of T-wave arrivals detected over the 4-year period. Yellow regions and dashed lines illustrate two examples of multi-array localization through azimuthal triangulation. Green stars mark the epicenters of two earthquakes (Mw 5.2, 2024-05-08 11:51:39 UTC, Vanuatu; and Mw 5.2, 2025-05-14 16:33:05 UTC, Pacific-Antarctic Ridge) corresponding to the observed T-waves.}
    \label{fig:T_localization}
\end{figure}

During the four-year SeaFOAM observation period, DASNet detected more than 83,000 T-wave events, far exceeding the number of cataloged earthquakes reported during the same interval (\Cref{fig:T_localization}a). Examination of the resulting detections reveals a diverse range of submarine activity, including major earthquakes and their aftershock sequences, prolonged episodes of volcanic unrest, and recurring local activity (\Cref{fig:T_localization}b). While many of these sequences coincide with independently documented seismic and volcanic events, others lack obvious counterparts in existing catalogs, suggesting that continuous DAS observations may capture a broader spectrum of submarine activity than is currently represented by conventional monitoring systems. 

To investigate the origins of these detections, we estimated the back azimuth of each T-wave by applying beamforming to 2–10 Hz bandpass-filtered data and retaining energy with apparent velocities near 1.5 km s$^{-1}$ through an f-k filter (Figure S4). The azimuth corresponding to the maximum beam power was taken as the most probable direction of arrival. The resulting four-year catalog reveals three persistent source regions (\Cref{fig:T_localization}a,c). The first cluster is characterized by predominantly southern back azimuths, consistent with sources along the Southeast Pacific spreading system. The second cluster exhibits southwestern back azimuths, indicating energy arriving from the Tonga–Vanuatu subduction zone. The third cluster is associated with western back azimuths, suggesting contributions from the Izu–Bonin–Mariana arc region. These back azimuths represent propagation directions rather than unique source locations, and the actual sources may lie anywhere along the corresponding great-circle paths. Notably, the southern cluster also contains numerous recurring unidentified T-wave sequences that lack obvious associations with cataloged large earthquakes. These signals exhibit pronounced seasonal variability and, in some cases, gradual azimuthal migration through time (\Cref{fig:T_localization}a,b). While distant seismic sources undoubtedly contribute to the southern cluster, the observed temporal patterns are difficult to explain solely by far-field earthquake activity and may instead indicate contributions from more local or regional submarine sources. Azimuthal constraints are weaker for southern arrivals because of the SeaFOAM cable geometry, leading to larger uncertainties that could be reduced through longer or more geometrically diverse DAS deployments (Figure S5). 
We also observe a relative paucity of detections near incidence offsets approaching 90$^{\circ}$, where arrivals propagate nearly perpendicular to the average cable orientation. This pattern likely reflects directional variations in array sensitivity associated with the cable geometry.

To further constrain source locations and verify azimuths identified by DASNet, we applied multi-array localization using simultaneous detections from SeaFOAM and two additional DAS arrays (\Cref{fig:T_localization}c, Figure S4 and Figure S6). We applied triangulation based on independent azimuthal estimates from separate cables to obtain constraints on remote source locations. In the first case, we used a 4-day overlapping deployment between the SeaFOAM DAS and the Ocean Observatories Initiative (OOI) Regional Cabled Array in May 2024, during which multiple coincident T-wave events were recorded \citep{romanowicz2023seafoam, lipovsky2024rapid, shi2024multiplexed}. A representative event on May 8, 2024 (observed at 13:38:05 UTC on OOI and 13:35:54 UTC on SeaFOAM) was jointly analyzed. The intersection of the two azimuthal estimates closely aligns with the epicenter of a Mw 5.2 earthquake that occurred offshore Vanuatu at 11:51:39 UTC. The predicted T-wave travel times agree with the observed arrivals on both DAS arrays, validating the multi-array localization method and confirming the southwestern Pacific as a T-wave source region. In the second case, we localized another coincident event recorded by SeaFOAM and the offshore Chile F1 DAS network operated by the French Seismological and Geodetic Network (RESIF) \citep{https://doi.org/10.15778/resif.f1}. On May 14, 2025, both arrays recorded a T-wave whose triangulated source location falls on the Pacific-Antarctic Ridge, coinciding with a Mw 5.2 earthquake at 16:33:05 UTC. Because the F1 DAS array data are provided at a 5-km spatial interval, the beamforming results are degraded due to spatial aliasing. Together, these examples demonstrate how distributed DAS arrays can move beyond detection and toward source attribution of ocean-basin-scale acoustic activity. As long-duration submarine deployments continue to expand, globally distributed DAS networks may provide a powerful framework for locating and monitoring seismic and volcanic activity across regions that remain sparsely instrumented by conventional observing systems, while potentially revealing previously unrecognized submarine activity.

\subsection{Tracking bioacoustic and anthropogenic sources} 
Vocalizations are central to the life history of many marine species, particularly baleen whales, which rely on low-frequency vocalizations for communication and navigation across vast oceanic distances \citep[]{payne1971songs, tyack2008implications}. Fin whales (\textit{Balaenoptera physalus}) and blue whales (\textit{Balaenoptera musculus}) produce particularly powerful and stereotyped calls \citep[]{mcdonald1995blue, roman2014whales, simon2010singing, ryan2022oceanic, oestreich2022acoustic}, making them well suited for passive acoustic monitoring (PAM). PAM has become a key tool for studying the spatial distribution and seasonal migration patterns of these species, especially in remote or visually inaccessible regions \citep{mcdonald1995blue, vsirovic2007blue, oestreich2024listening}. 
However, conventional PAM systems, such as hydrophone arrays, are often limited by sparse deployment and low spatial resolution \citep{ahonen2021interannual}. 
Recent studies have shown that DAS is capable of capturing whale vocalizations even in challenging environments and can be used for identifying and tracking whale behavior \citep{bouffaut2022eavesdropping, landro2022sensing, wilcock2023distributed, xiao2025imaging, goestchel2025enhancing}. 
DASNet's ability to automatically detect and classify diverse whale calls at scale enables long-term monitoring of seasonal distribution and behavior patterns across multi-year DAS deployments. We further leverage the model's capacity for vessel tracking to demonstrate a promising direction for quantifying anthropogenic noise exposure and its potential impact on marine ecosystems.

\begin{figure}[htbp]
    \centering
    \includegraphics[width=0.8\textwidth]{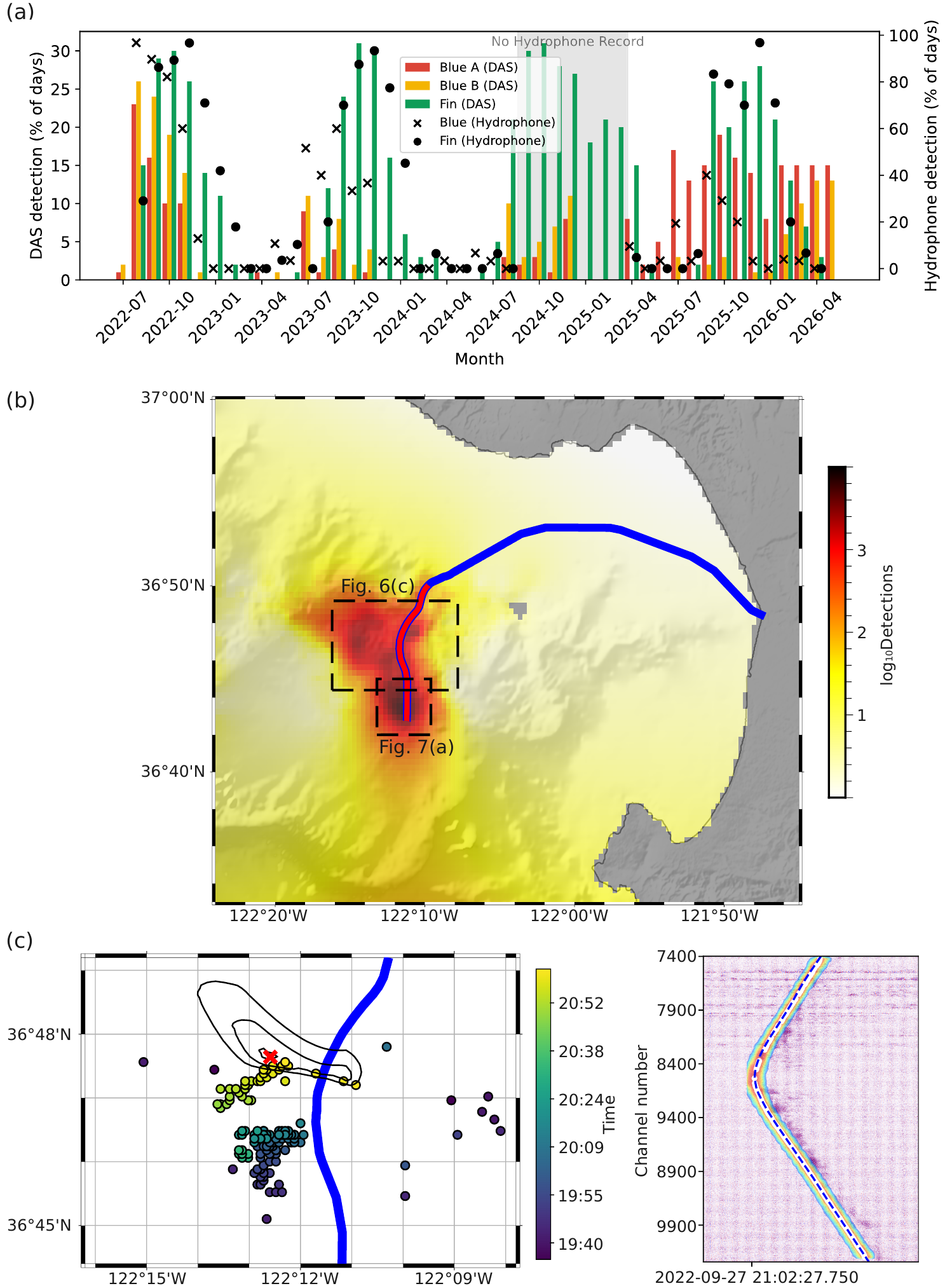}
    \caption{Whale call detection and localization. (a) Monthly occurrence of DAS-detected fin whale calls and blue whale A and B calls, reported as the percentage of days per month with detections. A day is counted as having a DAS detection if whale calls occupy $>$3$\%$ of the day; for the MBARI hydrophone catalog, a day is counted when the daily Call Index exceeds 1.05 (see Figure S7). (b) Spatial distribution of all fin whale calls localized over the four-year record. Results far from the cable should be interpreted with caution due to limitations of cable geometry and instrument sensitivity. (c) Case study spanning approximately 2~h. The left panel shows locations of 148 calls detected during this interval; the red cross marks the call whose waveform is shown at right. Black concentric contours denote misfit levels of 0.1~s, 0.2~s, and 0.3~s. In the right panel, the white curve traces arrival-time picks extracted from the model’s probabilistic mask, and the blue curve shows the predicted arrival-time curve computed from the estimated source location.}
    \label{fig:whale}
\end{figure}

During the monitoring period, the SeaFOAM DAS array recorded extensive vocalizations from fin and blue whales, including more than 420,000 fin whale calls and over 90,000 blue whale calls. The temporal distribution of these detections reveals distinct seasonal patterns consistent with known migratory behaviors (\Cref{fig:whale}a). Specifically, blue whale vocalizations usually appeared earlier in the season but over a shorter period, peaking between July and October, whereas fin whale call detections started later and persisted longer, spanning from August to the following April. In addition to seasonal variability, both species exhibited notable interannual variability in call activity, with fin whale detections increasing slightly across the record  
and blue whale detections declining after 2022 \citep{ryan2025audible}. 
To validate these seasonal and annual distributions, we compared DASNet detections with independent records from the Monterey Accelerated Research System (MARS) hydrophone (\Cref{fig:whale}a). Despite differences in sensing modality, spatial coverage, and detection criteria, the two systems yielded consistent seasonal peaks during overlapping time windows, supporting the robustness of DAS-based passive acoustic monitoring.

In addition to detecting whale vocalizations, DASNet provides precise arrival-time estimates for individual calls, enabling automatic source localization. We localized detected whale calls using a grid-based approach 
that minimizes the L1 misfit between theoretical and model-predicted arrival times \citep{xiao2025imaging}, assuming a nominal source depth of 15~m \citep{stimpert2015sound} and a constant water velocity (\Cref{fig:whale}c). 
Most localized calls cluster near the continental shelf slopes (\Cref{fig:whale}b and Figure S8), consistent with previous studies suggesting that such topographic features enhance biological productivity through upwelling, mixing, and nutrient retention, creating energetically favorable foraging habitats that attract baleen whales \citep{moors2014submarine}. Note that 
the spatial density of localizations is also influenced by the detection sensitivity of the DAS cable and the limitations of the localization algorithm, which provides weaker constraints in the direction perpendicular to the cable (\Cref{fig:whale}c). Therefore, the localization heatmap should be interpreted as a convolution of true biological activity and observation constraints. 
These results demonstrate that DAS infrastructure, paired with automatic detection capabilities of DASNet, offers a promising approach for large-scale, long-duration monitoring of marine mammal distribution.

\begin{figure}[htbp]
    \centering
    \includegraphics[width=0.75\textwidth]{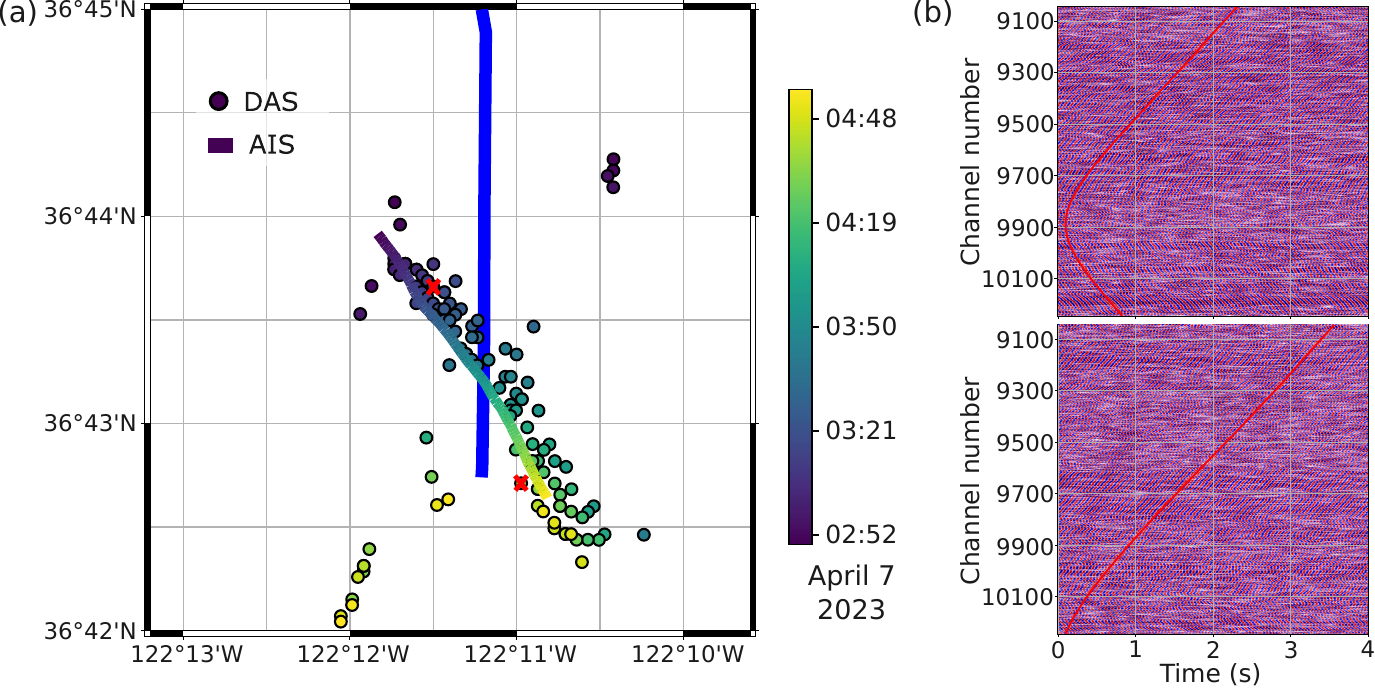}
    \caption{Vessel detection and localization. (a) Locations of vessel signals detected by DAS on April 7, 2023. Estimated source positions closely match AIS-reported vessel locations. Note that mirror-image solutions appear on both the east and west sides of the cable, reflecting the inherent left-right ambiguity of a near-linear array. The blue vertical line indicates the DAS cable. (b) Two signal segments corresponding to the red cross positions in (a), and the predicted arrival curves based on the localization results are marked with red curves.}
    \label{fig:vessel_locate}
\end{figure}

Besides ecological observations, DASNet can also support the monitoring of anthropogenic acoustic sources, such as vessel traffic, which poses growing threats to baleen whales through acoustic masking, habitat displacement, and ship strikes \citep{national2003ocean, mcdonald2006increases}. Although vessel noise typically lacks clear arrivals, DASNet detects these continuous acoustic signals through its bounding-box predictions. Combining these detections with a beamforming-inspired migration approach \citep{paap2025leveraging}, we can estimate the spatial trajectories of moving vessels. 
As a demonstration, we analyzed a two-hour DAS recording from April 7, 2023, during which a vessel with available Automatic Identification System (AIS) data passed near the SeaFOAM cable. DASNet detected 127 vessel signal segments, and the inferred locations traced a coherent trajectory that closely matches the vessel's movement as recorded by AIS (\Cref{fig:vessel_locate}). 
As with whale vocalizations, localization ambiguity inherent to the linear array geometry remains a challenge, producing mirror-image solutions on both sides of the cable. This limitation can be mitigated by deploying curved cable geometries or supplemental arrays. 
The ability to track vessels alongside whale calls positions DAS as a multi-purpose tool for integrated monitoring of marine ecosystems and the human activities that affect them.

\section{Discussion}

DAS provides a unique opportunity to observe the ocean as a coupled system over spatial scales from meters to tens of kilometers and temporal scales from seconds to seasons. Prior DAS studies commonly addressed individual processes in isolation, such as detecting earthquakes, tracking whale calls, or monitoring vessel traffic, requiring separate workflows with distinct preprocessing steps and manual inspection. In this work, we have developed 
DASNet, a unified instance-segmentation framework that reformulates DAS analysis around event-centric representations. Instead of interpreting continuous wavefields channel by channel, DASNet identifies each event as an individual spatiotemporal instance, predicting bounding boxes, arrival-time masks, and class labels for downstream scientific analyses. 
The unified framework enables a single DAS array to function as a multi-purpose observatory capable of distinguishing diverse signal sources. It supports local and regional earthquake detection and localization, identification and azimuthal characterization of distant T-wave sources, systematic monitoring and tracking of baleen whale vocalizations, and localization of nearby vessel traffic.

Reliable automated event detection and classification are central to interpreting large-scale DAS observations. We evaluated the performance of DASNet on datasets constructed through a semi-supervised refinement process. The results indicate that DASNet achieves consistent detection performance across well-represented signal classes and in continuous data. Comparisons with independent seismic catalogs, hydrophone-based whale call records, and vessel tracking data further support the validity of the detected events, demonstrating that the automated predictions are broadly consistent with established observational systems. These cross-validation results provide confidence that the extracted event catalogs capture physically meaningful signals rather than spurious detections.

Accurate arrival-time measurements are equally important for quantitative analyses such as event localization and subsurface imaging. Across the evaluated signal classes, DASNet achieves arrival-time precision adequate for the applications demonstrated in this study.
An important factor contributing to this performance is the two-stage detection-and-measurement strategy adopted by DASNet. Rather than directly predicting arrivals from continuous waveforms, the model first identifies candidate events and subsequently estimates arrival times within event-dependent temporal windows. This design allows the arrival-picking task to focus on a localized temporal context while maintaining sensitivity to signals spanning a wide range of durations. The temporal-anchor mechanism further refines this process by adaptively selecting the feature region used for arrival-time estimation. For earthquake analysis, this design helps to accommodate the large variation in signal duration between local and distant earthquakes. 
The picked seismic phases exhibit slightly reduced timing precision relative to impulsive acoustic signals. This discrepancy is consistent with the inherent difficulty of manually picking emergent phase onsets for lower-SNR events in DAS records, which introduces label variability that propagates into training and evaluation, ultimately limiting the achievable precision.

Operational deployment of event-centric DAS analysis further requires computational efficiency and stability on continuous data streams. 
Benchmarking on three months of continuous SeaFOAM data (2024-08 to 2024-10), one of the most event-dense periods in our observations, 
shows that the inference time per 60~s-segment rarely exceeds 0.5~s on a standard Google Cloud GPU instance with 8 vCPUs and an NVIDIA T4 GPU, which is 
sufficiently efficient for real-time monitoring of diverse seismic and acoustic events (Figure S9). Because DASNet employs a two-stage architecture (see Methods), the second-stage branches for classification, bounding-box regression, and mask generation require additional computation when the number of detected instances increases. However, individual 60~s segments in the SeaFOAM dataset rarely contain fifteen or more distinct signals, so the overall two-stage overhead remains limited under typical operating conditions.

Despite these capabilities, several challenges remain.
One limitation of DASNet is that the current training dataset does not yet encompass the full diversity of signal types present in continuous DAS recordings. 
As a result, certain signal classes that are either previously unrecognized or represent atypical variants of known types remain difficult to detect. Signals outside the training distribution may also be assigned to the most similar known classes by the model, because many signal types share common characteristics relative to background noise. 
One example is a category of signals that may represent candidate T-waves from the North Pacific (Figure S10). These signals often exhibit reduced coherence and less distinct moveout patterns compared with the more commonly observed T-waves, potentially reflecting generation in shallower water and enhanced scattering along complex propagation paths. 
Because their classification remains uncertain and representative examples were limited, these signals were not included in the current training set. However, an important advantage of the DASNet framework is its ability to support rapid iterative expansion through semi-supervised learning (see Methods). Once emerging or ambiguously classified signal types are identified and verified, a small number of representative examples can be incorporated into the training set and propagated through successive refinement cycles. This semi-supervised learning process provides an effective approach for continuously improving monitoring coverage as new signal types are recognized.

Another challenge is generalization across arrays. DASNet in this study was trained on SeaFOAM data, and performance can degrade when applied to cables with different coupling conditions, noise characteristics, or spatial sampling. Although the architecture supports inputs of varying sizes, direct application to substantially larger or smaller arrays without additional adaptation or fine-tuning has not been evaluated. Potential directions to enhance model generalization include data augmentation that simulates various DAS configurations, including spatial and temporal resampling, adding realistic noise, and composing new inputs through cropping and stitching \citep{zhu2020seismic}, as well as semi-supervised fine-tuning on the target datasets. 
We did not adopt a fully automated self-training loop in this study, because small errors in pseudo-labels can accumulate and bias the model, but recent work has shown that when the base model is sufficiently accurate, iterative self-training can approach full automation \citep{kirillov2023segment}. 
As DAS datasets continue to grow in volume and diversity, developing robust strategies for model adaptation across different cable configurations and geological settings will be critical for reliable and scalable DAS-based monitoring and analysis.

In conclusion, DASNet transforms distributed acoustic sensing from a set of task-specific pipelines into a unified event-level observatory capable of simultaneously monitoring seismic activity, marine ecosystems, and anthropogenic noise along the same fiber. It enables the simultaneous detection and analysis of diverse signals, including earthquakes, T-waves, whale vocalizations, and vessel activity, supporting long-term monitoring of multiple ocean processes on a single DAS array while maintaining the computational efficiency required for continuous real-time operation. These results demonstrate that combining DAS with deep learning provides scalable, high-resolution monitoring of marine environments using existing telecommunications infrastructure. In regions where conventional seismic and bioacoustic instruments remain sparse or absent, expanding submarine DAS coverage could substantially enhance observational capability. Although further developments are needed to ensure robust generalization across diverse DAS arrays, this study provides a foundation for large-scale ocean monitoring through the integration of fiber-optic sensing and deep learning.

\section*{Acknowledgments}
We thank Wenbo Wu and John Ryan for helpful discussions and suggestions on marine acoustic analysis. We thank Michael Manga for constructive discussions on volcanic acoustic activity. We thank Martin Karrenbach and Victor Yartsev of OptaSense Inc. for their assistance in installing and configuring the Monterey Bay distributed acoustic sensing equipment.

\clearpage
\bibliography{references}

@inproceedings{he2017mask,
  title={Mask r-cnn},
  author={He, Kaiming and Gkioxari, Georgia and Doll{\'a}r, Piotr and Girshick, Ross},
  booktitle={Proceedings of the IEEE international conference on computer vision},
  pages={2961--2969},
  year={2017}
}

@article{zhan2020distributed,
  title={Distributed acoustic sensing turns fiber-optic cables into sensitive seismic antennas},
  author={Zhan, Zhongwen},
  journal={Seismological Research Letters},
  volume={91},
  number={1},
  pages={1--15},
  year={2020},
  publisher={Seismological Society of America}
}

@article{lindsey2021fiber,
  title={Fiber-optic seismology},
  author={Lindsey, Nathaniel J and Martin, Eileen R},
  journal={Annual Review of Earth and Planetary Sciences},
  volume={49},
  number={1},
  pages={309--336},
  year={2021},
  publisher={Annual Reviews}
}

@article{williams2019distributed,
  title={Distributed sensing of microseisms and teleseisms with submarine dark fibers},
  author={Williams, Ethan F and Fern{\'a}ndez-Ruiz, Mar{\'\i}a R and Magalhaes, Regina and Vanthillo, Roel and Zhan, Zhongwen and Gonz{\'a}lez-Herr{\'a}ez, Miguel and Martins, Hugo F},
  journal={Nature communications},
  volume={10},
  number={1},
  pages={5778},
  year={2019},
  publisher={Nature Publishing Group UK London}
}

@article{shen2024ocean,
  title={Ocean bottom distributed acoustic sensing for oceanic seismicity detection and seismic ocean thermometry},
  author={Shen, Zhichao and Wu, Wenbo},
  journal={Journal of Geophysical Research: Solid Earth},
  volume={129},
  number={3},
  pages={e2023JB027799},
  year={2024},
  publisher={Wiley Online Library}
}

@article{wilcock2023distributed,
  title={Distributed acoustic sensing recordings of low-frequency whale calls and ship noise offshore Central Oregon},
  author={Wilcock, William SD and Abadi, Shima and Lipovsky, Bradley P},
  journal={JASA Express Letters},
  volume={3},
  number={2},
  year={2023},
  publisher={AIP Publishing}
}

@article{xiao2024detection,
  title={Detection of earthquake infragravity and tsunami waves with underwater distributed acoustic sensing},
  author={Xiao, Han and Spica, Zack J and Li, Jiaxuan and Zhan, Zhongwen},
  journal={Geophysical Research Letters},
  volume={51},
  number={2},
  pages={e2023GL106767},
  year={2024},
  publisher={Wiley Online Library}
}

@article{lior2022imaging,
  title={Imaging an underwater basin and its resonance modes using optical fiber distributed acoustic sensing},
  author={Lior, Itzhak and Mercerat, E Diego and Rivet, Diane and Sladen, Anthony and Ampuero, Jean-Paul},
  journal={Seismological Society of America},
  volume={93},
  number={3},
  pages={1573--1584},
  year={2022}
}

@article{lior2021detection,
  title={On the detection capabilities of underwater distributed acoustic sensing},
  author={Lior, Itzhak and Sladen, Anthony and Rivet, Diane and Ampuero, Jean-Paul and Hello, Yann and Becerril, Carlos and Martins, Hugo F and Lamare, Patrick and Jestin, Camille and Tsagkli, Stavroula and others},
  journal={Journal of Geophysical Research: Solid Earth},
  volume={126},
  number={3},
  pages={e2020JB020925},
  year={2021},
  publisher={Wiley Online Library}
}

@article{sladen2019distributed,
  title={Distributed sensing of earthquakes and ocean-solid Earth interactions on seafloor telecom cables},
  author={Sladen, Anthony and Rivet, Diane and Ampuero, Jean Paul and De Barros, Louis and Hello, Yann and Calbris, Ga{\"e}tan and Lamare, Patrick},
  journal={Nature communications},
  volume={10},
  number={1},
  pages={5777},
  year={2019},
  publisher={Nature Publishing Group UK London}
}

@article{williams2022surface,
  title={Surface gravity wave interferometry and ocean current monitoring with ocean-bottom DAS},
  author={Williams, Ethan F and Zhan, Zhongwen and Martins, Hugo F and Fern{\'a}ndez-Ruiz, Mar{\'\i}a R and Mart{\'\i}n-L{\'o}pez, Sonia and Gonz{\'a}lez-Herr{\'a}ez, Miguel and Callies, J{\"o}rn},
  journal={Journal of Geophysical Research: Oceans},
  volume={127},
  number={5},
  pages={e2021JC018375},
  year={2022},
  publisher={Wiley Online Library}
}

@article{romanowicz2023seafoam,
  title={SeaFOAM: A year-long DAS deployment in Monterey Bay, California},
  author={Romanowicz, Barbara and Allen, Richard and Brekke, Knute and Chen, Li-Wei and Gou, Yuancong and Henson, Ivan and Marty, Julien and Neuhauser, Doug and Pardini, Brian and Taira, Taka’aki and others},
  journal={Seismological Research Letters},
  volume={94},
  number={5},
  pages={2348--2359},
  year={2023},
  publisher={Seismological Society of America}
}

@article{shi2024multiplexed,
  title={Multiplexed Distributed Acoustic Sensing offshore Central Oregon},
  author={Shi, Qibin and Williams, Ethan F and Lipovsky, Bradley P and Denolle, Marine A and Wilcock, William SD and Kelley, Deborah S and Schoedl, Katelyn},
  journal={Authorea Preprints},
  year={2024},
  publisher={Authorea}
}

@article{yu2024daseventnet,
  title={DASEventNet: AI-based microseismic detection on distributed acoustic sensing data from the Utah FORGE well 16A (78)-32 hydraulic stimulation},
  author={Yu, Pengliang and Zhu, Tieyuan and Marone, Chris and Elsworth, Derek and Yu, Mingzhao},
  journal={Journal of Geophysical Research: Solid Earth},
  volume={129},
  number={9},
  pages={e2024JB029102},
  year={2024},
  publisher={Wiley Online Library}
}

@article{zhu2023seismic,
  title={Seismic arrival-time picking on distributed acoustic sensing data using semi-supervised learning},
  author={Zhu, Weiqiang and Biondi, Ettore and Li, Jiaxuan and Yin, Jiuxun and Ross, Zachary E and Zhan, Zhongwen},
  journal={Nature Communications},
  volume={14},
  number={1},
  pages={8192},
  year={2023},
  publisher={Nature Publishing Group UK London}
}

@article{zhu2019phasenet,
  title={PhaseNet: a deep-neural-network-based seismic arrival-time picking method},
  author={Zhu, Weiqiang and Beroza, Gregory C},
  journal={Geophysical Journal International},
  volume={216},
  number={1},
  pages={261--273},
  year={2019},
  publisher={Oxford University Press}
}

@article{mousavi2020earthquake,
  title={Earthquake transformer—an attentive deep-learning model for simultaneous earthquake detection and phase picking},
  author={Mousavi, S Mostafa and Ellsworth, William L and Zhu, Weiqiang and Chuang, Lindsay Y and Beroza, Gregory C},
  journal={Nature communications},
  volume={11},
  number={1},
  pages={3952},
  year={2020},
  publisher={Nature Publishing Group UK London}
}

@article{landro2022sensing,
  title={Sensing whales, storms, ships and earthquakes using an Arctic fibre optic cable},
  author={Landr{\o}, Martin and Bouffaut, L{\'e}a and Kriesell, Hannah Joy and Potter, John Robert and R{\o}rstadbotnen, Robin Andr{\'e} and Taweesintananon, Kittinat and Johansen, St{\aa}le Emil and Brenne, Jan Kristoffer and Haukanes, Aksel and Schjelderup, Olaf and others},
  journal={Scientific Reports},
  volume={12},
  number={1},
  pages={19226},
  year={2022},
  publisher={Nature Publishing Group UK London}
}

@article{bouffaut2022eavesdropping,
  title={Eavesdropping at the speed of light: Distributed acoustic sensing of baleen whales in the Arctic},
  author={Bouffaut, L{\'e}a and Taweesintananon, Kittinat and Kriesell, Hannah J and R{\o}rstadbotnen, Robin A and Potter, John R and Landr{\o}, Martin and Johansen, St{\aa}le E and Brenne, Jan K and Haukanes, Aksel and Schjelderup, Olaf and others},
  journal={Frontiers in Marine Science},
  volume={9},
  pages={901348},
  year={2022},
  publisher={Frontiers}
}

@article{wang2018ground,
  title={Ground motion response to an ML 4.3 earthquake using co-located distributed acoustic sensing and seismometer arrays},
  author={Wang, Herbert F and Zeng, Xiangfang and Miller, Douglas E and Fratta, Dante and Feigl, Kurt L and Thurber, Clifford H and Mellors, Robert J},
  journal={Geophysical Journal International},
  volume={213},
  number={3},
  pages={2020--2036},
  year={2018},
  publisher={Oxford University Press}
}

@article{lindsey2020broadband,
  title={On the broadband instrument response of fiber-optic DAS arrays},
  author={Lindsey, Nathaniel J and Rademacher, Horst and Ajo-Franklin, Jonathan B},
  journal={Journal of Geophysical Research: Solid Earth},
  volume={125},
  number={2},
  pages={e2019JB018145},
  year={2020},
  publisher={Wiley Online Library}
}

@article{muir2022wavefield,
  title={Wavefield-based evaluation of DAS instrument response and array design},
  author={Muir, Jack B and Zhan, Zhongwen},
  journal={Geophysical Journal International},
  volume={229},
  number={1},
  pages={21--34},
  year={2022},
  publisher={Oxford University Press}
}

@article{zhu2022earthquake,
  title={Earthquake phase association using a Bayesian Gaussian mixture model},
  author={Zhu, Weiqiang and McBrearty, Ian W and Mousavi, S Mostafa and Ellsworth, William L and Beroza, Gregory C},
  journal={Journal of Geophysical Research: Solid Earth},
  volume={127},
  number={5},
  pages={e2021JB023249},
  year={2022},
  publisher={Wiley Online Library}
}

@article{ross2019phaselink,
  title={PhaseLink: A deep learning approach to seismic phase association},
  author={Ross, Zachary E and Yue, Yisong and Meier, Men-Andrin and Hauksson, Egill and Heaton, Thomas H},
  journal={Journal of Geophysical Research: Solid Earth},
  volume={124},
  number={1},
  pages={856--869},
  year={2019},
  publisher={Wiley Online Library}
}

@article{zhu2019seismic,
  title={Seismic signal denoising and decomposition using deep neural networks},
  author={Zhu, Weiqiang and Mousavi, S Mostafa and Beroza, Gregory C},
  journal={IEEE Transactions on Geoscience and Remote Sensing},
  volume={57},
  number={11},
  pages={9476--9488},
  year={2019},
  publisher={IEEE}
}

@article{zhang2019rapid,
  title={Rapid earthquake association and location},
  author={Zhang, Miao and Ellsworth, William L and Beroza, Gregory C},
  journal={Seismological Research Letters},
  volume={90},
  number={6},
  pages={2276--2284},
  year={2019},
  publisher={GeoScienceWorld}
}

@inproceedings{he2016deep,
  title={Deep residual learning for image recognition},
  author={He, Kaiming and Zhang, Xiangyu and Ren, Shaoqing and Sun, Jian},
  booktitle={Proceedings of the IEEE conference on computer vision and pattern recognition},
  pages={770--778},
  year={2016}
}

@inproceedings{liu2021swin,
  title={Swin transformer: Hierarchical vision transformer using shifted windows},
  author={Liu, Ze and Lin, Yutong and Cao, Yue and Hu, Han and Wei, Yixuan and Zhang, Zheng and Lin, Stephen and Guo, Baining},
  booktitle={Proceedings of the IEEE/CVF international conference on computer vision},
  pages={10012--10022},
  year={2021}
}

@inproceedings{liu2022convnet,
  title={A convnet for the 2020s},
  author={Liu, Zhuang and Mao, Hanzi and Wu, Chao-Yuan and Feichtenhofer, Christoph and Darrell, Trevor and Xie, Saining},
  booktitle={Proceedings of the IEEE/CVF conference on computer vision and pattern recognition},
  pages={11976--11986},
  year={2022}
}

@article{tolstoy1950t,
  title={The T phase of shallow-focus earthquakes},
  author={Tolstoy, Ivan and Ewing, Maurice},
  journal={Bulletin of the Seismological Society of America},
  volume={40},
  number={1},
  pages={25--51},
  year={1950},
  publisher={The Seismological Society of America}
}

@article{paap2025leveraging,
  title={Leveraging Distributed Acoustic Sensing for monitoring vessels using submarine fiber-optic cables},
  author={Paap, Bob and Vandeweijer, Vincent and van Wees, Jan-Diederik and Kraaijpoel, Dirk},
  journal={Applied Ocean Research},
  volume={154},
  pages={104422},
  year={2025},
  publisher={Elsevier}
}

@article{wiggins2005blue,
  title={Blue whale (Balaenoptera musculus) diel call patterns offshore of Southern California},
  author={Wiggins, Sean M and Oleson, Erin M and McDonald, Mark A and Hildebrand, John A},
  journal={Aquatic Mammals},
  volume={31},
  number={2},
  pages={161},
  year={2005},
  publisher={Aquatic Mammals}
}

@article{wuestefeld2024global,
  title={The global DAS month of February 2023},
  author={Wuestefeld, Andreas and Spica, Zack J and Aderhold, Kasey and Huang, Hsin-Hua and Ma, Kuo-Fong and Lai, Voon Hui and Miller, Meghan and Urmantseva, Lena and Zapf, Daniel and Bowden, Daniel C and others},
  journal={Seismological Research Letters},
  volume={95},
  number={3},
  pages={1569--1577},
  year={2024},
  publisher={Seismological Society of America}
}

@article{wu2020seismic,
  title={Seismic ocean thermometry},
  author={Wu, Wenbo and Zhan, Zhongwen and Peng, Shirui and Ni, Sidao and Callies, J{\"o}rn},
  journal={Science},
  volume={369},
  number={6510},
  pages={1510--1515},
  year={2020},
  publisher={American Association for the Advancement of Science}
}

@incollection{zhu2020seismic,
  title={Seismic signal augmentation to improve generalization of deep neural networks},
  author={Zhu, Weiqiang and Mousavi, S Mostafa and Beroza, Gregory C},
  booktitle={Advances in geophysics},
  volume={61},
  pages={151--177},
  year={2020},
  publisher={Elsevier}
}

@article{huot2022detection,
  title={Detection and characterization of microseismic events from fiber-optic DAS data using deep learning},
  author={Huot, Fantine and Lellouch, Ariel and Given, Paige and Luo, Bin and Clapp, Robert G and Nemeth, Tamas and Nihei, Kurt T and Biondi, Biondo L},
  journal={Seismological Society of America},
  volume={93},
  number={5},
  pages={2543--2553},
  year={2022}
}

@article{ma2023machine,
  title={Machine learning-assisted processing workflow for multi-fiber DAS microseismic data},
  author={Ma, Yuanyuan and Eaton, David and Igonin, Nadine and Wang, Chaoyi},
  journal={Frontiers in Earth Science},
  volume={11},
  pages={1096212},
  year={2023},
  publisher={Frontiers Media SA}
}

@inproceedings{kirillov2023segment,
  title={Segment anything},
  author={Kirillov, Alexander and Mintun, Eric and Ravi, Nikhila and Mao, Hanzi and Rolland, Chloe and Gustafson, Laura and Xiao, Tete and Whitehead, Spencer and Berg, Alexander C and Lo, Wan-Yen and others},
  booktitle={Proceedings of the IEEE/CVF international conference on computer vision},
  pages={4015--4026},
  year={2023}
}

@article{carvajal2022worldwide,
  title={Worldwide signature of the 2022 Tonga volcanic tsunami},
  author={Carvajal, Mat{\'\i}as and Sep{\'u}lveda, Ignacio and Gubler, Alejandra and Garreaud, Ren{\'e}},
  journal={Geophysical Research Letters},
  volume={49},
  number={6},
  pages={e2022GL098153},
  year={2022},
  publisher={Wiley Online Library}
}

@article{harbitz2006mechanisms,
  title={Mechanisms of tsunami generation by submarine landslides: a short review.},
  author={Harbitz, Carl B and L{\o}vholt, Finn and Pedersen, Geir and Masson, Doug G},
  journal={Norwegian Journal of Geology/Norsk Geologisk Forening},
  volume={86},
  number={3},
  year={2006}
}

@article{kanamori1972mechanism,
  title={Mechanism of tsunami earthquakes},
  author={Kanamori, Hiroo},
  journal={Physics of the earth and planetary interiors},
  volume={6},
  number={5},
  pages={346--359},
  year={1972},
  publisher={Elsevier}
}

@article{omira2022global,
  title={Global Tonga tsunami explained by a fast-moving atmospheric source},
  author={Omira, R and Ramalho, RS and Kim, Jinyoung and Gonz{\'a}lez, Pablo J and Kadri, U and Miranda, JM and Carrilho, F and Baptista, MA},
  journal={Nature},
  volume={609},
  number={7928},
  pages={734--740},
  year={2022},
  publisher={Nature Publishing Group UK London}
}

@article{tepp2021seismo,
  title={The seismo-acoustics of submarine volcanic eruptions},
  author={Tepp, Gabrielle and Dziak, Robert P},
  journal={Journal of Geophysical Research: Solid Earth},
  volume={126},
  number={4},
  pages={e2020JB020912},
  year={2021},
  publisher={Wiley Online Library}
}

@article{hildebrand2009anthropogenic,
  title={Anthropogenic and natural sources of ambient noise in the ocean},
  author={Hildebrand, John A},
  journal={Marine Ecology Progress Series},
  volume={395},
  pages={5--20},
  year={2009}
}

@article{roman2014whales,
  title={Whales as marine ecosystem engineers},
  author={Roman, Joe and Estes, James A and Morissette, Lyne and Smith, Craig and Costa, Daniel and McCarthy, James and Nation, J Brian and Nicol, Stephen and Pershing, Andrew and Smetacek, Victor},
  journal={Frontiers in Ecology and the Environment},
  volume={12},
  number={7},
  pages={377--385},
  year={2014},
  publisher={Wiley Online Library}
}

@article{simon2010singing,
  title={Singing behavior of fin whales in the Davis Strait with implications for mating, migration and foraging},
  author={Simon, Malene and Stafford, Kathleen M and Beedholm, Kristian and Lee, Craig M and Madsen, Peter T},
  journal={The Journal of the Acoustical Society of America},
  volume={128},
  number={5},
  pages={3200--3210},
  year={2010},
  publisher={AIP Publishing}
}

@article{ryan2022oceanic,
  title={Oceanic giants dance to atmospheric rhythms: Ephemeral wind-driven resource tracking by blue whales},
  author={Ryan, John P and Benoit-Bird, Kelly J and Oestreich, William K and Leary, Paul and Smith, Kevin B and Waluk, Chad M and Cade, David E and Fahlbusch, James A and Southall, Brandon L and Joseph, John E and others},
  journal={Ecology letters},
  volume={25},
  number={11},
  pages={2435--2447},
  year={2022},
  publisher={Wiley Online Library}
}

@article{oestreich2022acoustic,
  title={Acoustic signature reveals blue whales tune life-history transitions to oceanographic conditions},
  author={Oestreich, William K and Abrahms, Briana and McKenna, Megan F and Goldbogen, Jeremy A and Crowder, Larry B and Ryan, John P},
  journal={Functional Ecology},
  volume={36},
  number={4},
  pages={882--895},
  year={2022},
  publisher={Wiley Online Library}
}

@article{toomey2014cascadia,
  title={The Cascadia Initiative: A sea change in seismological studies of subduction zones},
  author={Toomey, Douglas R and Allen, Richard M and Barclay, Andrew H and Bell, Samuel W and Bromirski, Peter D and Carlson, Richard L and Chen, Xiaowei and Collins, John A and Dziak, Robert P and Evers, Brent and others},
  journal={Oceanography},
  volume={27},
  number={2},
  pages={138--150},
  year={2014},
  publisher={JSTOR}
}

@article{suetsugu2014broadband,
  title={Broadband ocean-bottom seismology},
  author={Suetsugu, Daisuke and Shiobara, Hajime},
  journal={Annual Review of Earth and Planetary Sciences},
  volume={42},
  number={1},
  pages={27--43},
  year={2014},
  publisher={Annual Reviews}
}

@article{lindsey2019illuminating,
  title={Illuminating seafloor faults and ocean dynamics with dark fiber distributed acoustic sensing},
  author={Lindsey, Nathaniel J and Dawe, T Craig and Ajo-Franklin, Jonathan B},
  journal={Science},
  volume={366},
  number={6469},
  pages={1103--1107},
  year={2019},
  publisher={American Association for the Advancement of Science}
}

@article{fernandez2022seismic,
  title={Seismic monitoring with distributed acoustic sensing from the near-surface to the deep oceans},
  author={Fern{\'a}ndez-Ruiz, Mar{\'\i}a R and Martins, Hugo F and Williams, Ethan F and Becerril, Carlos and Magalh{\~a}es, Regina and Costa, Luis and Martin-Lopez, Sonia and Jia, Zhensheng and Zhan, Zhongwen and Gonz{\'a}lez-Herr{\'a}ez, Miguel},
  journal={Journal of Lightwave Technology},
  volume={40},
  number={5},
  pages={1453--1463},
  year={2022},
  publisher={OSA}
}

@article{fox2001monitoring,
  title={Monitoring Pacific Ocean seismicity from an autonomous hydrophone array},
  author={Fox, Christopher G and Matsumoto, Haruyoshi and Lau, Tai-Kwan Andy},
  journal={Journal of Geophysical Research: Solid Earth},
  volume={106},
  number={B3},
  pages={4183--4206},
  year={2001},
  publisher={Wiley Online Library}
}

@article{smith2002hydroacoustic,
  title={Hydroacoustic monitoring of seismicity at the slow-spreading Mid-Atlantic Ridge},
  author={Smith, Deborah K and Tolstoy, Maya and Fox, Christopher G and Bohnenstiehl, DelWayne R and Matsumoto, Haru and J. Fowler, Matthew},
  journal={Geophysical Research Letters},
  volume={29},
  number={11},
  pages={13--1},
  year={2002},
  publisher={Wiley Online Library}
}

@article{wech2018using,
  title={Using earthquakes, T waves, and infrasound to investigate the eruption of Bogoslof volcano, Alaska},
  author={Wech, Aaron and Tepp, Gabrielle and Lyons, John and Haney, Matt},
  journal={Geophysical Research Letters},
  volume={45},
  number={14},
  pages={6918--6925},
  year={2018},
  publisher={Wiley Online Library}
}

@article{xiao2025imaging,
  title={Imaging underwater faults and tracking whales with optical fiber sensing},
  author={Xiao, Han and Zhang, Shane and Moss, Robb and Zhan, Zhongwen},
  journal={Seismological Research Letters},
  volume={96},
  number={2A},
  pages={678--690},
  year={2025},
  publisher={Seismological Society of America}
}

@article{oestreich2024listening,
  title={Listening to animal behavior to understand changing ecosystems},
  author={Oestreich, William K and Oliver, Ruth Y and Chapman, Melissa S and Go, Madeline and McKenna, Megan F},
  journal={Trends in Ecology \& Evolution},
  year={2024},
  publisher={Elsevier}
}

@article{vsirovic2007blue,
  title={Blue and fin whale call source levels and propagation range in the Southern Ocean},
  author={{\v{S}}irovi{\'c}, Ana and Hildebrand, John A and Wiggins, Sean M},
  journal={The Journal of the Acoustical Society of America},
  volume={122},
  number={2},
  pages={1208--1215},
  year={2007},
  publisher={AIP Publishing}
}

@article{mcdonald1995blue,
  title={Blue and fin whales observed on a seafloor array in the Northeast Pacific},
  author={McDonald, Mark A and Hildebrand, John A and Webb, Spahr C},
  journal={The Journal of the Acoustical Society of America},
  volume={98},
  number={2},
  pages={712--721},
  year={1995},
  publisher={Acoustical Society of America}
}

@article{ahonen2021interannual,
  title={Interannual variability in acoustic detection of blue and fin whale calls in the Northeast Atlantic High Arctic between 2008 and 2018},
  author={Ahonen, Heidi and Stafford, Kathleen M and Lydersen, Christian and Berchok, Catherine L and Moore, Sue E and Kovacs, Kit M},
  journal={Endangered Species Research},
  volume={45},
  pages={209--224},
  year={2021}
}

@article{stimpert2015sound,
  title={Sound production and associated behavior of tagged fin whales (Balaenoptera physalus) in the Southern California Bight},
  author={Stimpert, Alison K and DeRuiter, Stacy L and Falcone, Erin A and Joseph, John and Douglas, Annie B and Moretti, David J and Friedlaender, Ari S and Calambokidis, John and Gailey, Glenn and Tyack, Peter L and others},
  journal={Animal Biotelemetry},
  volume={3},
  number={1},
  pages={23},
  year={2015},
  publisher={Springer}
}

@article{yin2023earthquake,
  title={Earthquake magnitude with DAS: A transferable data-based scaling relation},
  author={Yin, Jiuxun and Zhu, Weiqiang and Li, Jiaxuan and Biondi, Ettore and Miao, Yaolin and Spica, Zack J and Viens, Lo{\"\i}c and Shinohara, Masanao and Ide, Satoshi and Mochizuki, Kimihiro and others},
  journal={Geophysical Research Letters},
  volume={50},
  number={10},
  pages={e2023GL103045},
  year={2023},
  publisher={Wiley Online Library}
}

@misc{https://doi.org/10.15778/resif.f1,
  doi = {10.15778/RESIF.F1},
  author = {Rivet, Diane and Baillet, Marie and Trabattoni, Alister and Van Den Ende, Martijn and Chèze, Jérôme and Maron, Christophe and Sanchez, Rodrigo and Barrientos, Sergio and {Epos-France}},
  language = {en},
  title = {Three underwater fiber-optic cables offshore of central Chile},
  publisher = {Epos-France Seismological Data Centre},
  year = {2025},
  copyright = {Creative Commons Attribution 4.0 International}
}

@article{lipovsky2024rapid,
  title={RAPID: Multiplexed distributed acoustic sensing (DAS) at the ocean observatory initiative (OOI) regional cabled array (RCA)},
  author={Lipovsky, Bradley P},
  journal={NSF Award Number 2415521. Directorate for Geosciences},
  volume={24},
  number={2415521},
  pages={15521},
  year={2024}
}

@article{moors2014submarine,
  title={Submarine canyons as important habitat for cetaceans, with special reference to the Gully: a review},
  author={Moors-Murphy, Hilary B},
  journal={Deep Sea Research Part II: Topical Studies in Oceanography},
  volume={104},
  pages={6--19},
  year={2014},
  publisher={Elsevier}
}

@article{ryan2025audible,
  title={Audible changes in marine trophic ecology: Baleen whale song tracks foraging conditions in the eastern North Pacific},
  author={Ryan, John P and Oestreich, William K and Benoit-Bird, Kelly J and Waluk, Chad M and Rueda, Carlos A and Cline, Danelle E and Zhang, Yanwu and Cheeseman, Ted and Calambokidis, John and Fahlbusch, James A and others},
  journal={Plos one},
  volume={20},
  number={2},
  pages={e0318624},
  year={2025},
  publisher={Public Library of Science San Francisco, CA USA}
}

@article{gou2025leveraging,
  title={Leveraging submarine DAS arrays for offshore earthquake early warning: A case study in Monterey Bay, California},
  author={Gou, Yuancong and Allen, Richard M and Zhu, Weiqiang and Taira, Taka’aki and Chen, Li-Wei},
  journal={Bulletin of the Seismological Society of America},
  volume={115},
  number={2},
  pages={516--532},
  year={2025},
  publisher={Seismological Society of America}
}

@article{ding2025dasformer,
  title={DASFormer: self-supervised pretraining for earthquake monitoring},
  author={Ding, Qianggang and Shen, Zhichao and Zhu, Weiqiang and Liu, Bang},
  journal={Visual Intelligence},
  volume={3},
  number={1},
  pages={14},
  year={2025},
  publisher={Springer}
}

@article{van2021self,
  title={A self-supervised deep learning approach for blind denoising and waveform coherence enhancement in distributed acoustic sensing data},
  author={Van den Ende, Martijn and Lior, Itzhak and Ampuero, Jean-Paul and Sladen, Anthony and Ferrari, Andr{\'e} and Richard, C{\'e}dric},
  journal={IEEE Transactions on Neural Networks and Learning Systems},
  volume={34},
  number={7},
  pages={3371--3384},
  year={2021},
  publisher={IEEE}
}

@article{goestchel2025enhancing,
  title={Enhancing fin whale vocalizations in distributed acoustic sensing data},
  author={Goestchel, Quentin and Wilcock, William SD and Abadi, Shima},
  journal={The Journal of the Acoustical Society of America},
  volume={157},
  number={5},
  pages={3655--3666},
  year={2025},
  publisher={AIP Publishing}
}

@article{dosovitskiy2020image,
  title={An image is worth 16x16 words: Transformers for image recognition at scale},
  author={Dosovitskiy, Alexey},
  journal={arXiv preprint arXiv:2010.11929},
  year={2020}
}

@article{tyack2008implications,
  title={Implications for marine mammals of large-scale changes in the marine acoustic environment},
  author={Tyack, Peter L},
  journal={Journal of Mammalogy},
  volume={89},
  number={3},
  pages={549--558},
  year={2008},
  publisher={American Society of Mammalogists Allen Marketing \& Management, 810 East 10th~…}
}

@article{payne1971songs,
  title={Songs of Humpback Whales: Humpbacks emit sounds in long, predictable patterns ranging over frequencies audible to humans.},
  author={Payne, Roger S and McVay, Scott},
  journal={Science},
  volume={173},
  number={3997},
  pages={585--597},
  year={1971},
  publisher={American Association for the Advancement of Science}
}

@article{van2021evaluating,
  title={Evaluating seismic beamforming capabilities of distributed acoustic sensing arrays},
  author={van den Ende, Martijn PA and Ampuero, Jean-Paul},
  journal={Solid Earth},
  volume={12},
  number={4},
  pages={915--934},
  year={2021},
  publisher={Copernicus GmbH}
}

@article{mcdonald2006increases,
  title={Increases in deep ocean ambient noise in the Northeast Pacific west of San Nicolas Island, California},
  author={McDonald, Mark A and Hildebrand, John A and Wiggins, Sean M},
  journal={The Journal of the Acoustical Society of America},
  volume={120},
  number={2},
  pages={711--718},
  year={2006},
  publisher={AIP Publishing}
}

@article{national2003ocean,
  title={Ocean noise and marine mammals},
  author={National Research Council and Division on Earth and Life Studies and Ocean Studies Board and Committee on Potential Impacts of Ambient Noise in the Ocean on Marine Mammals},
  year={2003},
  publisher={National Academies Press}
}

@article{10.1093/gji/ggag061,
    author = {Xiao, Han and Tilmann, Frederik and van den Ende, Martijn and Rivet, Diane and Loureiro, Afonso and Tsuji, Takeshi and Ugalde, Arantza and Shi, Qibin and Denolle, Marine A},
    title = {DeepSubDAS: an earthquake phase picker from submarine distributed acoustic sensing data},
    journal = {Geophysical Journal International},
    volume = {245},
    number = {2},
    pages = {ggag061},
    year = {2026},
    month = {02},
    issn = {1365-246X},
    doi = {10.1093/gji/ggag061},
    url = {https://doi.org/10.1093/gji/ggag061},
    eprint = {https://academic.oup.com/gji/article-pdf/245/2/ggag061/66817494/ggag061.pdf},
}

@article{shen2026unsupervised,
  title={Unsupervised characterization of rain-induced seismic noise in urban fiber-optic networks using deep embedded clustering},
  author={Shen, Junzhu and Zhu, Tieyuan},
  journal={Water Resources Research},
  volume={62},
  number={1},
  pages={e2025WR041137},
  year={2026},
  publisher={Wiley Online Library}
}

@misc{GVP_Ahyi_2023,
  author       = {{Global Volcanism Program}},
  title        = {Report on Ahyi (United States)},
  editor       = {Sennert, S.},
  year         = {2023},
  howpublished = {Weekly Volcanic Activity Report, 8 March--14 March 2023},
  institution  = {Smithsonian Institution and U.S. Geological Survey},
  url          = {https://volcano.si.edu/showreport.cfm?gvpvar=GVP.WVAR20230308-284141},
  note         = {Accessed: 2026-06-17}
}

@article{xi2024deep,
  title={Deep learning for deep earthquakes: insights from OBS observations of the Tonga subduction zone},
  author={Xi, Ziyi and Wei, S Shawn and Zhu, Weiqiang and Beroza, Gregory C and Jie, Yaqi and Saloor, Nooshin},
  journal={Geophysical Journal International},
  volume={238},
  number={2},
  pages={1073--1088},
  year={2024},
  publisher={Oxford University Press}
}

\end{document}


\maketitle

\noindent\textbf{This PDF file includes:}\\
Text S1\\
Figs. S1 to S10

\section*{}
\begin{figure}[H]
    \centering
    \includegraphics[width=0.95\textwidth]{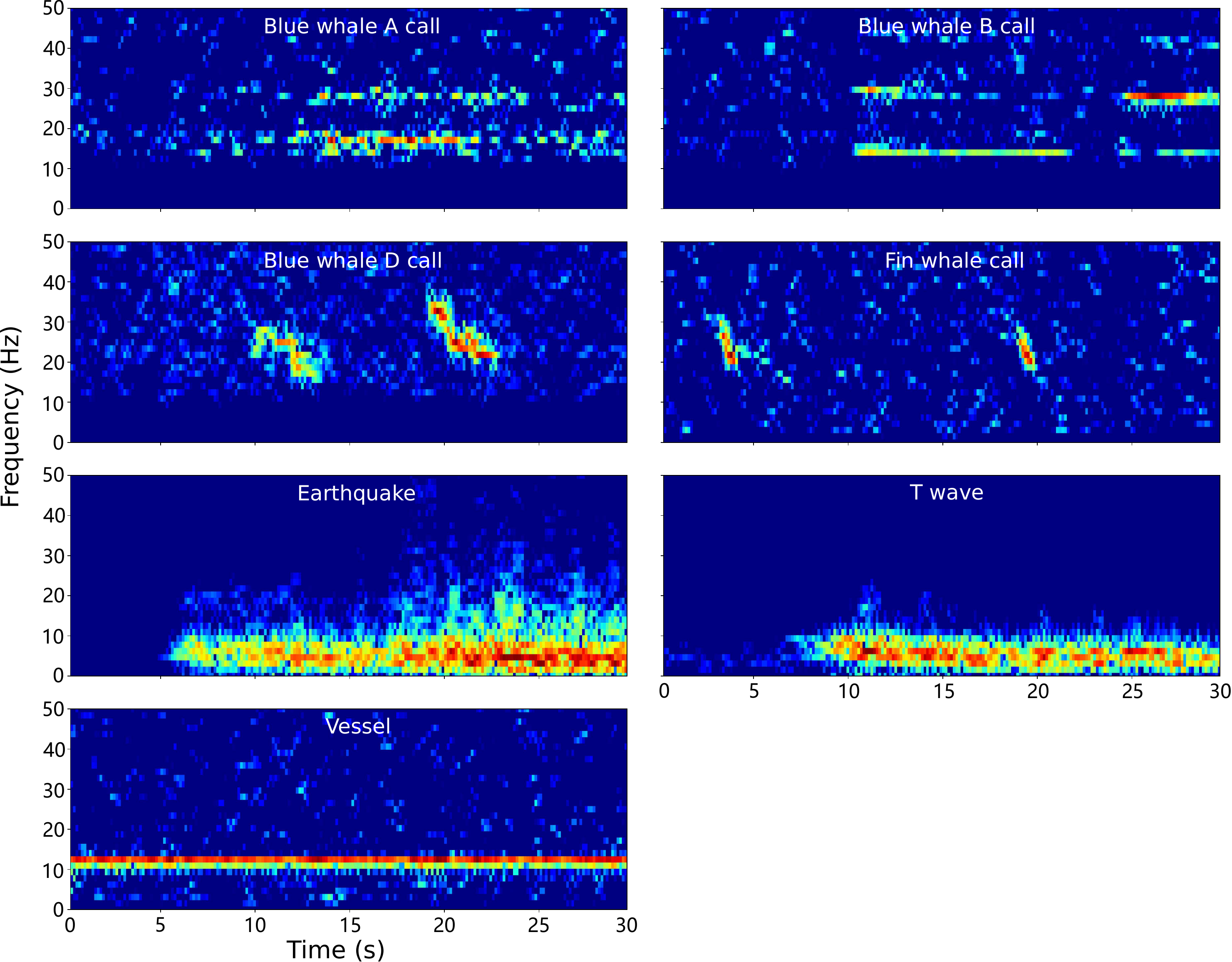}
    \caption{Spectrograms of representative examples for all signal types detected in the SeaFOAM dataset. Each panel shows the spectrogram of a single selected channel from the DAS recording. For whale vocalizations and vessel noise, a 10~Hz high-pass filter was applied, whereas a 2~Hz high-pass filter was used for the other signal types to suppress low-frequency background noise. The displayed spectral amplitudes were limited to a fixed dynamic range to enhance signal visibility.}
    \label{fig:spe}
\end{figure}

\begin{figure}[H]
    \centering
    \includegraphics[width=0.95\textwidth]{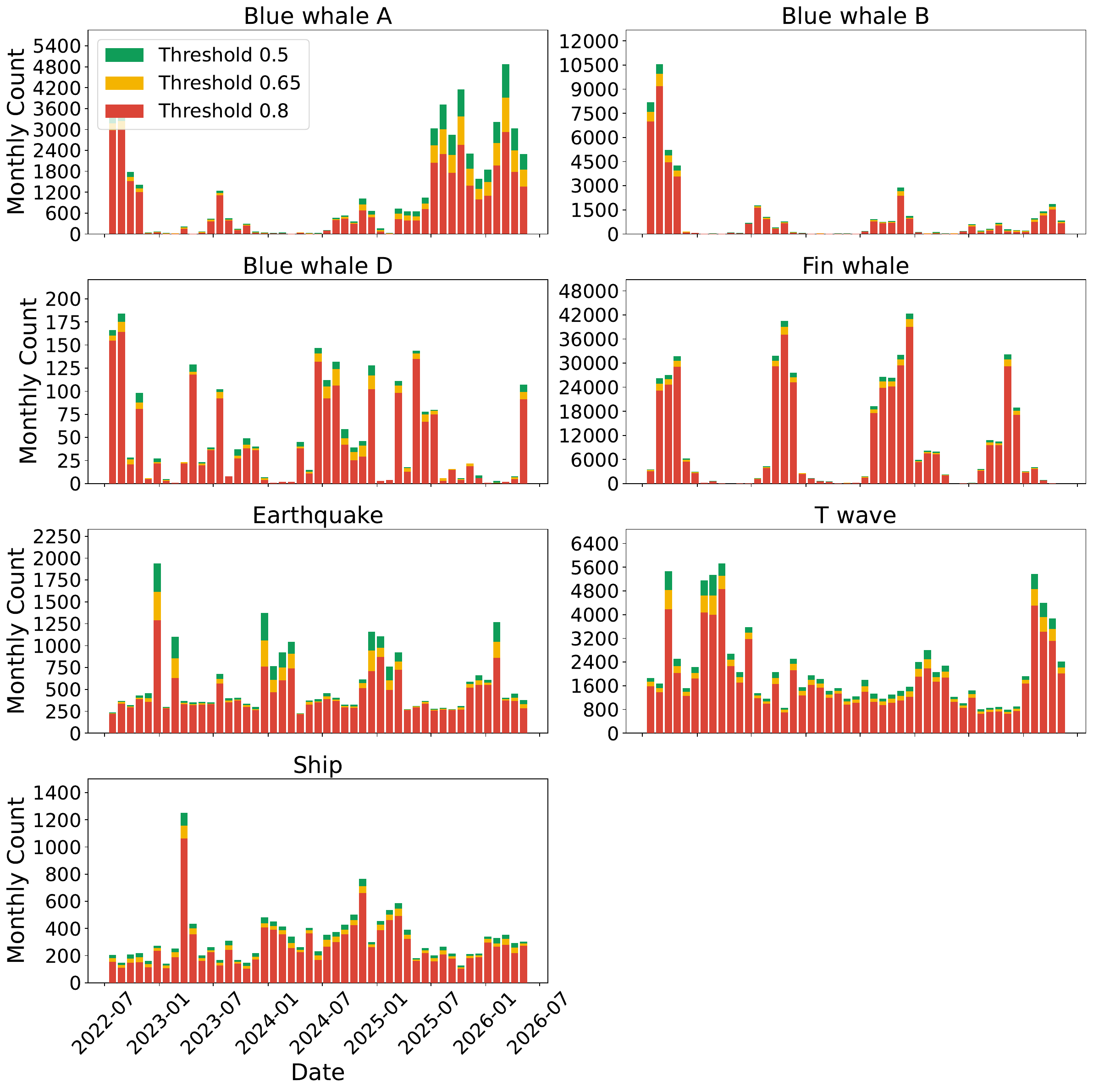}
    \caption{Comparison of DASNet detections across nearly 4 years of continuous data under different score thresholds. Detection counts at lower thresholds co-vary temporally with high-confidence detections, indicating that low-score detections are predominantly associated with real signal activity rather than spurious false positives.}
    \label{fig:diff_thre_comp}
\end{figure}

\begin{figure}[H]
    \centering
    \includegraphics[width=0.8\textwidth]{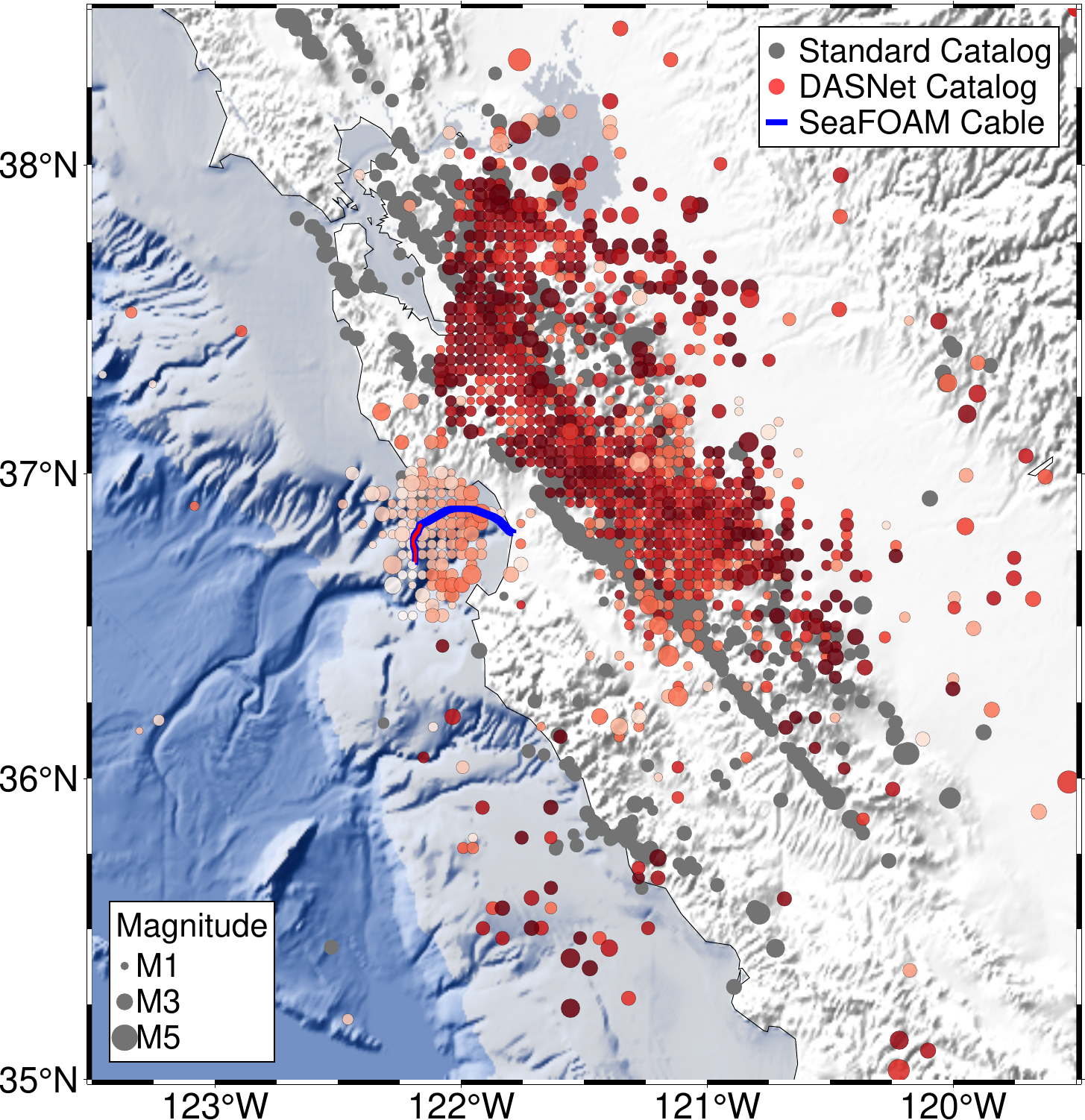}
    \caption{Earthquake locations estimated using P- and S-phase arrival times from the DAS array alone. The DAS-only solutions exhibit increased location uncertainty, especially for events farther from the cable, due to the limited azimuthal coverage of a single linear array and the reduced completeness of model-picked phase arrivals. This geometry provides weaker constraints on the back-azimuth and epicentral distance, resulting in larger lateral scatter in the inferred epicenters.}
    \label{fig:eq_location}
\end{figure}

\begin{figure}[H]
    \centering
    \includegraphics[width=0.8\textwidth]{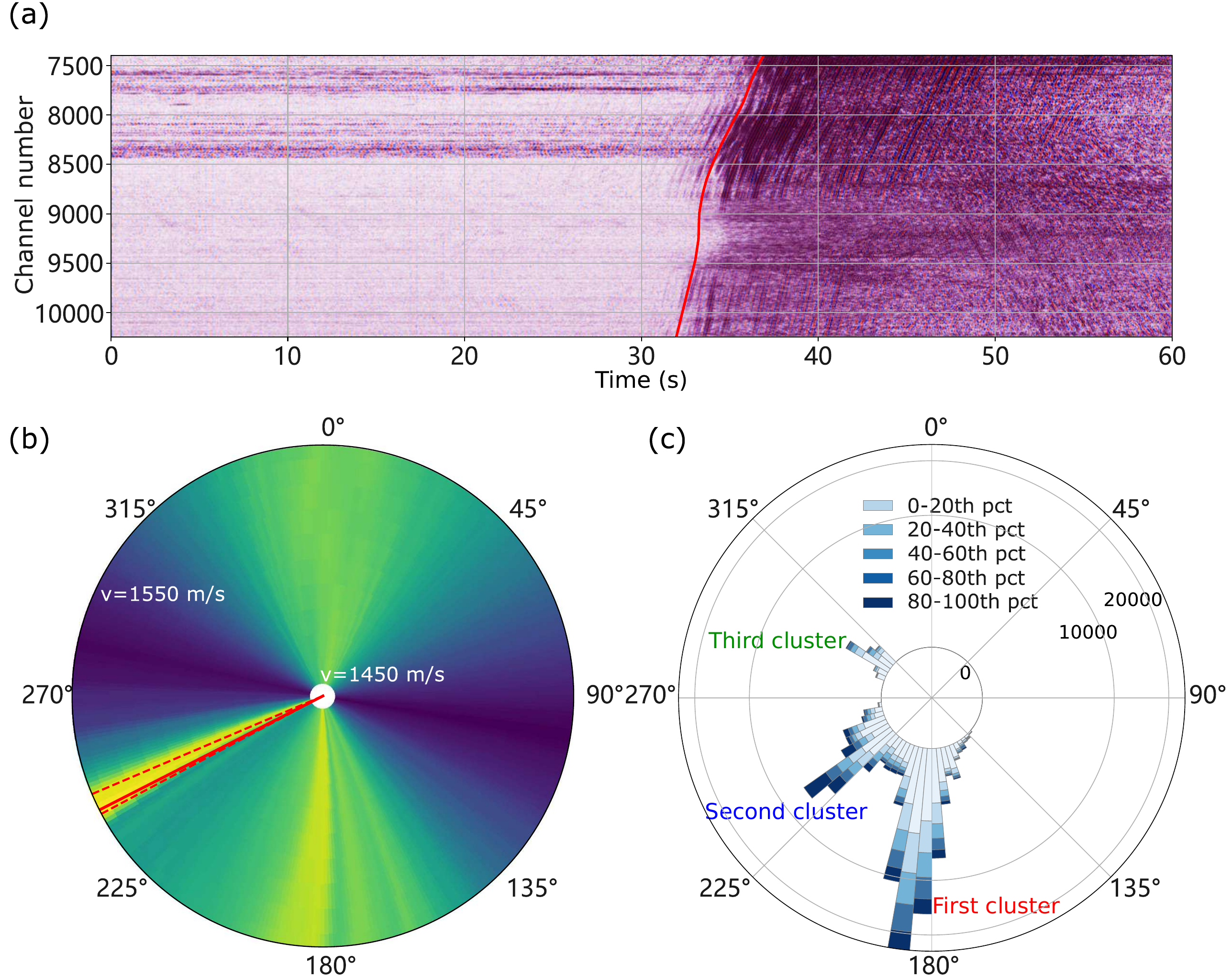}
    \caption{(a) T-wave waveforms recorded by the SeaFOAM cable, beginning at 2024-05-08 13:35:22 UTC. The red line indicates the theoretical arrival times computed from the estimated T-wave azimuth shown in (b) for the Vanuatu event (assuming a propagation velocity of 1.46~km$/$s, which provides the best fit to the observed signal onset), showing close agreement with the observed waveforms. (b) Heat maps derived via beamforming for azimuthal estimation. (c) Distribution of estimated T-wave source azimuths. The radial axis uses a square-root scale to improve visibility of lower-count azimuth bins. Color shading within each azimuth bin indicates the distribution of signal amplitudes, with darker colors representing stronger arrivals. Three prominent azimuthal clusters are observed.}
    \label{fig:T_wave_beamform}
\end{figure}

\begin{figure}[H]
    \centering
    \includegraphics[width=0.6\textwidth]{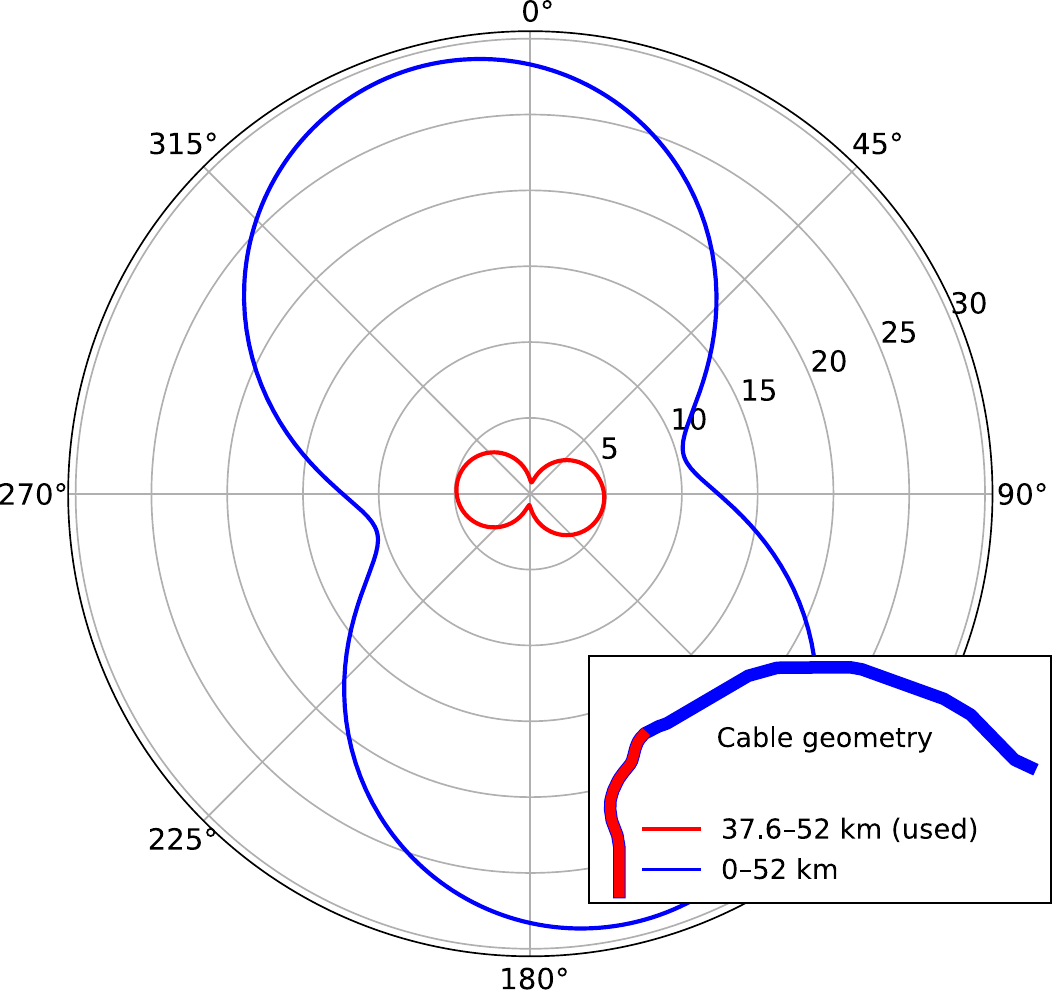}
    \caption{Directional sensitivity of T-wave beamforming. Sensitivity is quantified by measuring how strongly the predicted arrival-time curves across the DAS array change in response to small perturbations in propagation azimuth. The figure compares two cable geometries: the analyzed cable segment and the full cable. Azimuthal constraints are weakest for sources aligned with the cable axis, i.e., along-axis directions yield small inter-channel delay differences. Longer and more curved cable geometries markedly improve azimuthal resolution.}
    \label{fig:beam_form}
\end{figure}

\begin{figure}[H]
    \centering
    \includegraphics[width=0.95\textwidth]{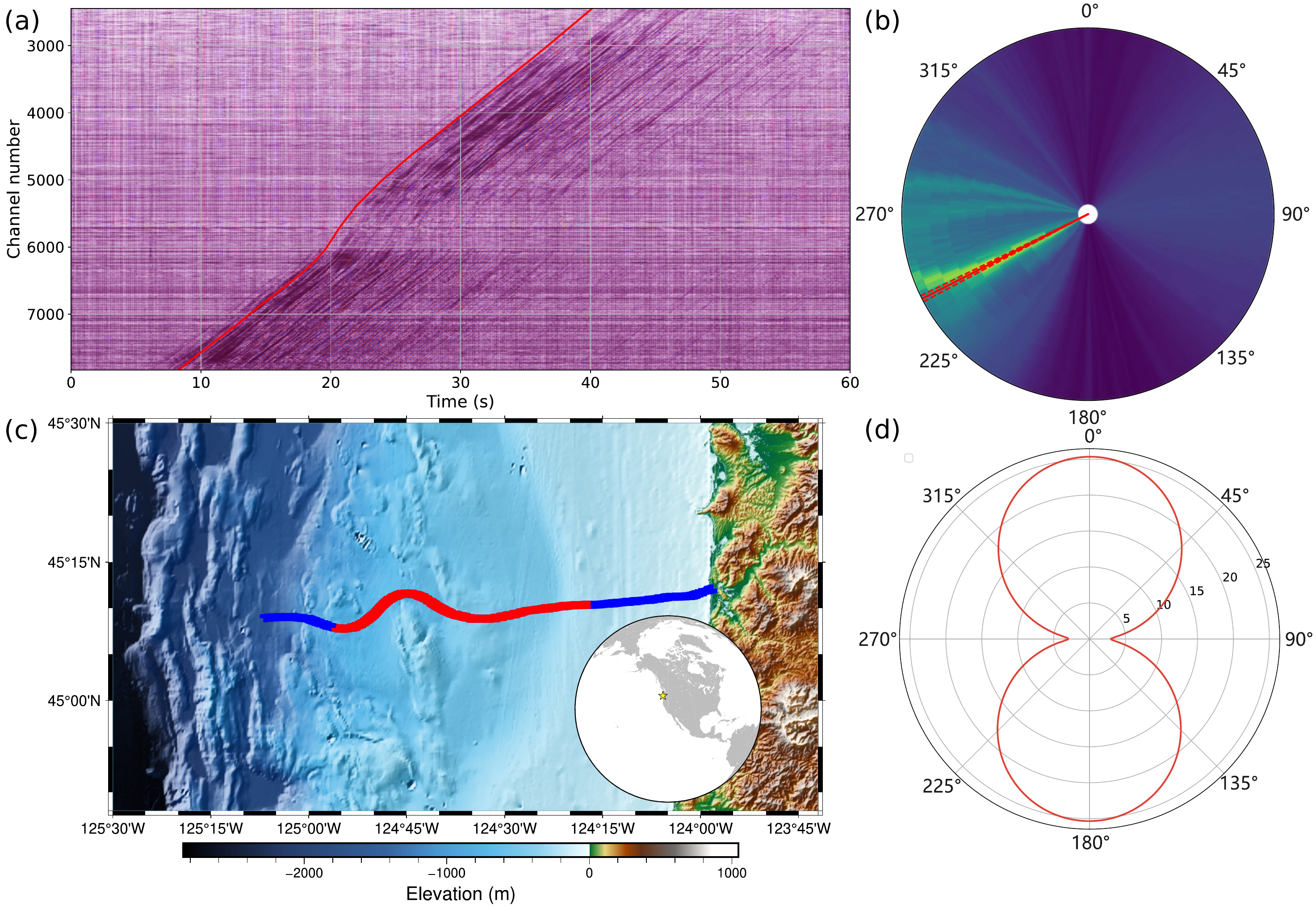}
    \caption{T-wave detection and azimuth estimation using the OOI Regional Cabled Array. (a) T-wave waveforms recorded by the OOI cable, beginning at 2024-05-08 13:38:05 UTC. The red lines show the theoretical time difference of arrival computed from the estimated T-wave azimuth in (b). The time offset has been removed. (b) Heatmap derived from beamforming for azimuth estimation. (c) Location and geometry of the OOI cable. (d) Azimuthal sensitivity of the OOI array.
    }
    \label{fig:OOI_Twave}
\end{figure}

\begin{figure}[H]
    \centering
    \includegraphics[width=0.95\textwidth]{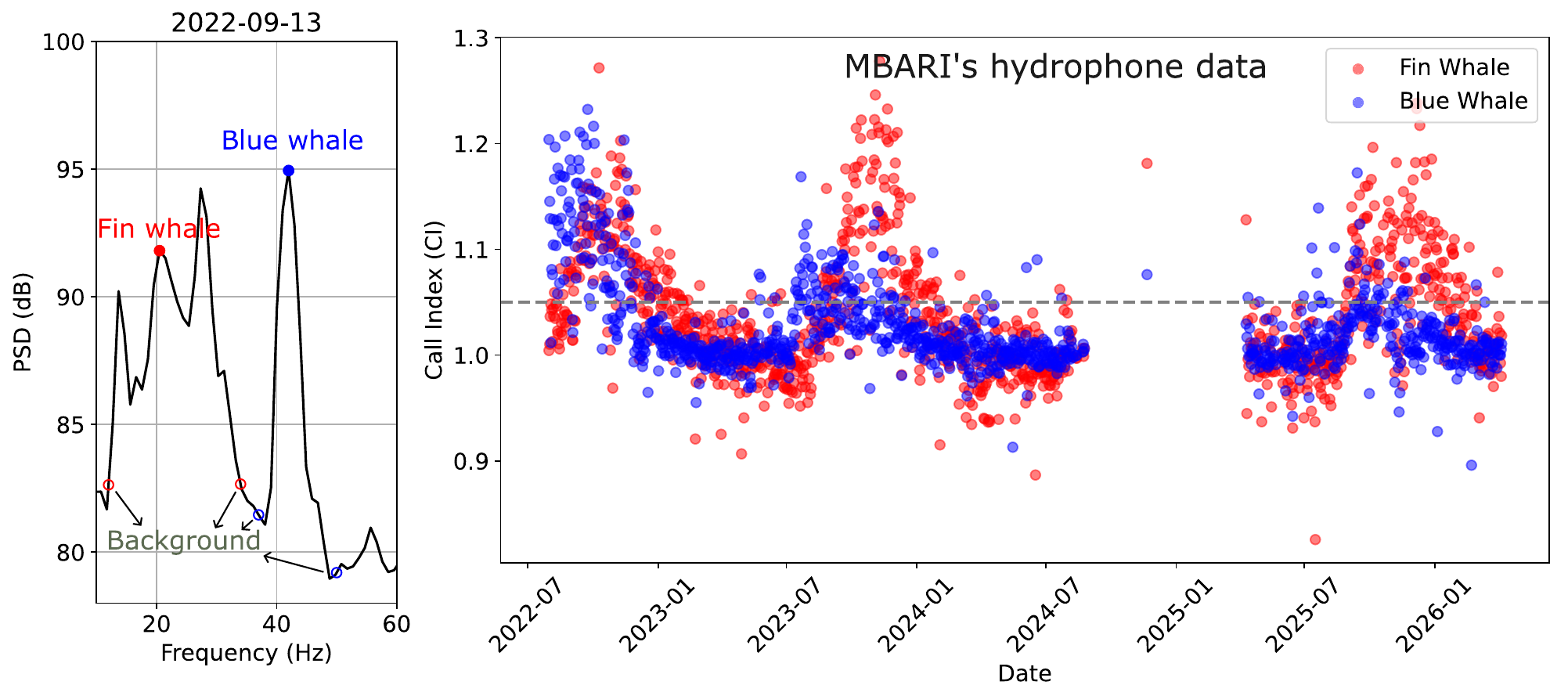}
    \caption{Call Index method for whale acoustic detection from the MBARI hydrophone. (a) Mean spectrum levels measured on 2022-09-13, showing peaks associated with fin whale 20~Hz pulses and the third harmonic of blue whale B calls (around 40~Hz). We compute the Call Index as the ratio of energy at the characteristic frequencies (solid dots) to energy at the background-noise frequencies (open dots). (b) Daily Call Index values. The gray dashed line marks the detection threshold (1.05) used to classify a day as whale-present.}
    \label{fig:mbari_call_index}
\end{figure}

\begin{figure}[H]
    \centering
    \includegraphics[width=0.95\textwidth]{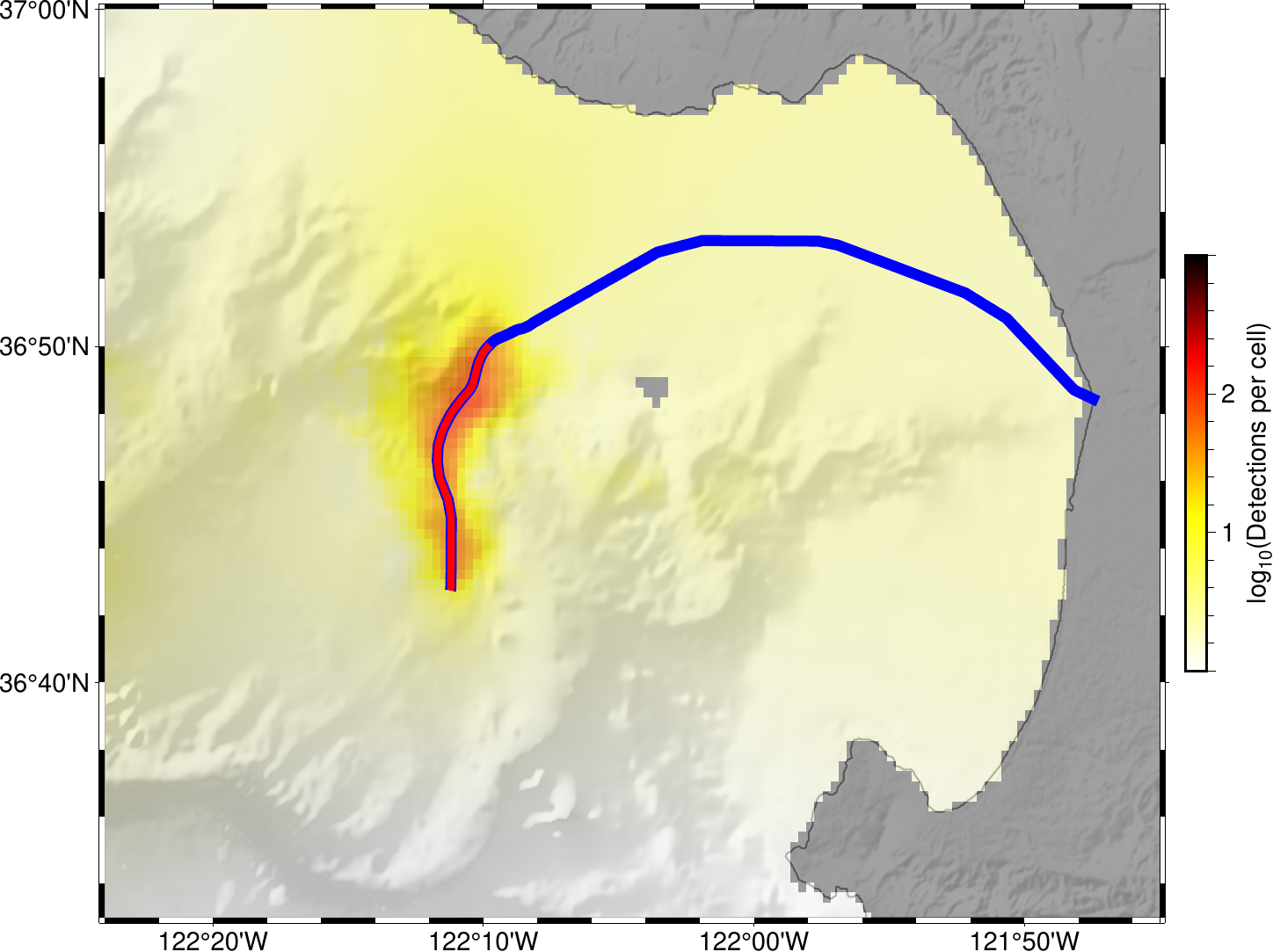}
    \caption{Spatial distribution of blue whale calls over the nearly four-year record. Compared to fin whales (Figure 4b in the main text), blue whales produce fewer calls.
    }
    \label{fig:bluewhale_heatmap}
\end{figure}

\begin{figure}[H]
    \centering
    \includegraphics[width=0.8\textwidth]{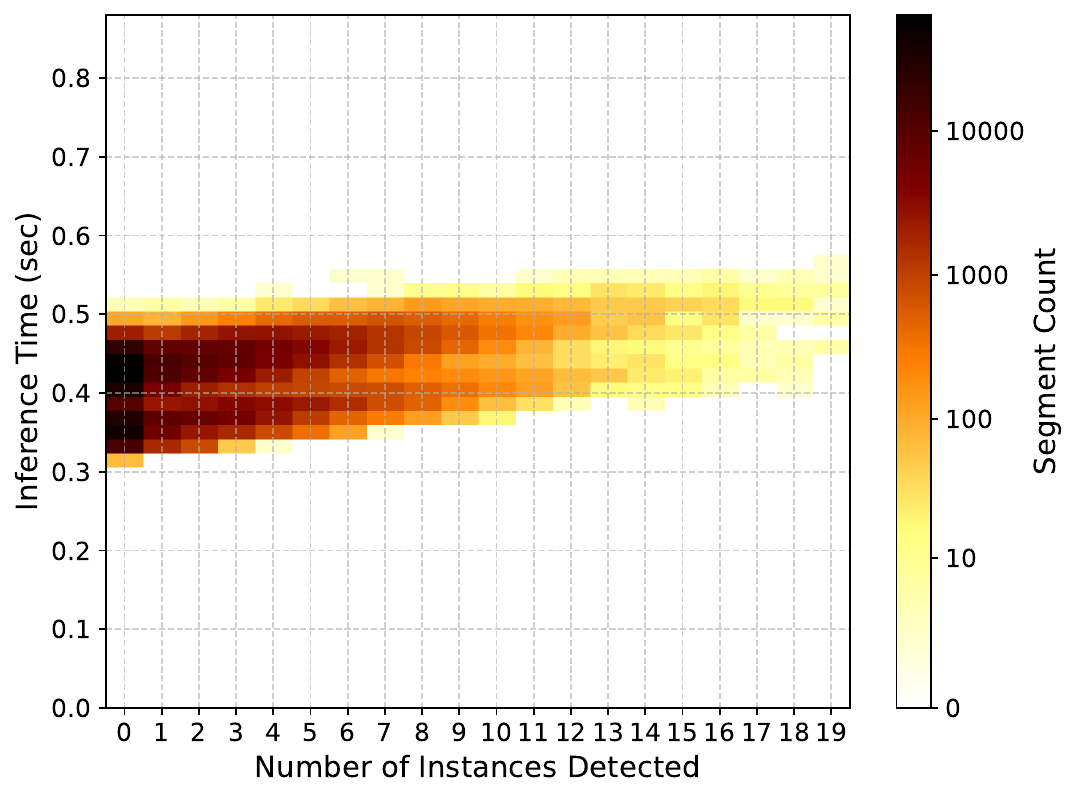}
    \caption{Inference time statistics for DASNet. Each input segment contains 2,845 channels and 12,000 time samples (60~s at 200~Hz). The x-axis shows the number of region proposals per segment that pass the first-stage filtering and are processed by the second-stage detection heads, and the y-axis shows the corresponding inference time. Inference time increases modestly with instance count but remains consistently low, supporting real-time monitoring applications.}
    \label{fig:inf_speed}
\end{figure}

\begin{figure}[H]
    \centering
    \includegraphics[width=0.95\textwidth]{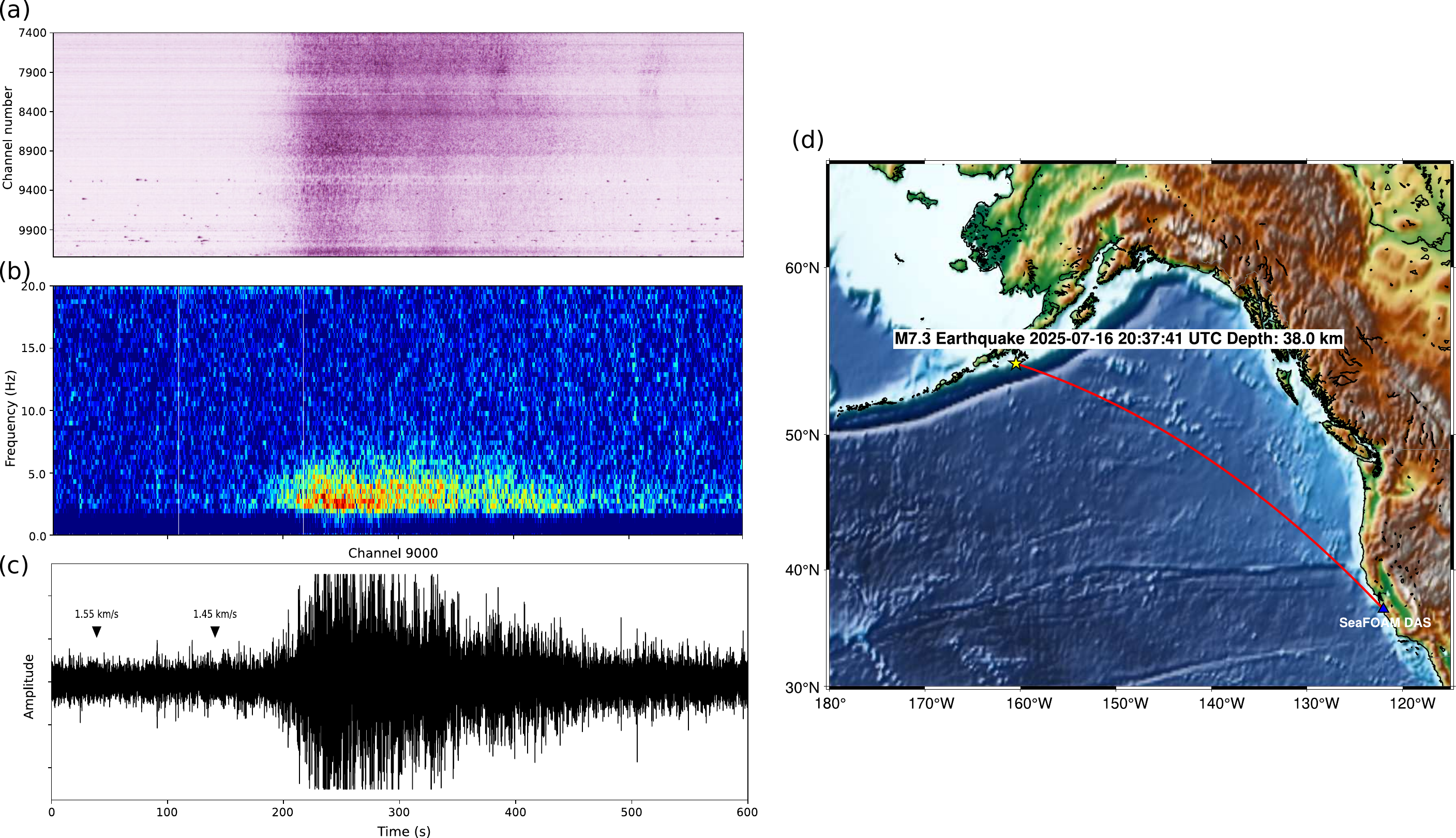}
    \caption{Example of a candidate T-wave from the North Pacific.
    (a) Spatio-temporal representations for the recorded signal.
    (b) Spectrogram for Channel 9,000, highlighting enhanced energy primarily in the 2-10 Hz band during the signal interval.
    (c) 2-10 Hz bandpass-filtered waveform for Channel 9,000, with inverted triangles indicating predicted arrival times corresponding to assumed propagation velocities of 1.55~km/s and 1.45~km/s.
    (d) Map view showing the epicenter of the M7.3 earthquake on 16 July 2025 and the great-circle path (red curve) to the SeaFOAM DAS array.}
    \label{fig:possible_t}
\end{figure}

The Supplementary Materials provide additional analyses and validation results that support the main conclusions of this study. 
It includes a detailed description of the DASNet architecture (Text~S1). 
We also provide spectrograms of representative channel records for each major signal type, corresponding to the examples shown in the main text (\Cref{fig:spe}). 
Additional comparisons of DASNet detections under different confidence thresholds are presented in \Cref{fig:diff_thre_comp}. 
Further earthquake location results based only on DAS phase arrivals are shown in \Cref{fig:eq_location}.  We also include representative examples of T-wave waveforms, beamforming results, and azimuth estimation (\Cref{fig:T_wave_beamform}), together with extended analyses of array geometry effects on T-wave azimuthal resolution (\Cref{fig:beam_form}). 
We also present examples from independent DAS and hydrophone datasets, including T-wave observations from the OOI Regional Cabled Array (\Cref{fig:OOI_Twave}), the whale Call Index (CI) from the MBARI hydrophone (\Cref{fig:mbari_call_index}), and blue whale detections from the SeaFOAM DAS array (\Cref{fig:bluewhale_heatmap}), demonstrating the generality of the proposed approach. 
Inference-time statistics (\Cref{fig:inf_speed}) are further provided to evaluate the computational efficiency of DASNet under varying detection scenarios. 
Finally, we show an example of a poorly structured T-wave-like signal from the high-latitude North Pacific (\Cref{fig:possible_t}), illustrating a type of signal variability that is not yet fully represented in the current training set.

\newpage
\noindent\textbf{Text S1. DASNet architecture}
\label{text:architecture}

DASNet performs joint detection and instance segmentation on two-dimensional DAS data (time $\times$ channel). 
The model predicts bounding boxes, class labels, and per-instance masks, from which signal arrival times can be estimated for each channel. 
The architecture consists of several stages described below.

\paragraph{Backbone and feature pyramid network.}
A ResNet-50 backbone is used to extract hierarchical features from the input DAS image. 
Feature maps from four backbone stages (with 256, 512, 1024, and 2048 channels, respectively) are fed into a Feature Pyramid Network (FPN), which produces multi-scale feature maps with 128 channels each. 
These shared features are used by the Region Proposal Network (RPN) as well as the box and mask heads.

\paragraph{Region proposal network.}
The RPN operates on each FPN level to generate candidate regions of interest. 
At each spatial location, it predicts objectness scores and bounding-box offsets for multiple anchors with aspect ratios of 0.5, 1.0, and 2.0. 
Non-maximum suppression (NMS) is applied to remove redundant proposals, and a fixed number of candidate regions are retained and passed to subsequent detection branches.

\paragraph{Temporal anchor network.}
Before mask prediction, each detection proposal is assigned a time-centered mask window to address the large spatial-temporal aspect ratio of DAS signals.
A Temporal Anchor Network takes the shared $7\times7$ RoI features (from the same multi-scale RoIAlign used by the box head) and predicts one temporal anchor per proposal by two fully connected layers (512 hidden units, ReLU).
The anchor is placed along the proposal time axis as
$y_{\mathrm{anchor}} = y_1 + f\,(y_2-y_1)$, where $(y_1,y_2)$ denote the proposal time bounds and $f\in[0,1]$.
The mask RoI keeps the same channel extent as the detection box and sets the time extent to
$[y_{\mathrm{anchor}} - b_{\mathrm{before}}L,\; y_{\mathrm{anchor}} + b_{\mathrm{after}}L]$,
where
$L = N_{\mathrm{ch}} / \alpha$ is the number of spatial channels in the DAS input segment,
and $\alpha = v_{\mathrm{ref}}\,\Delta t / \Delta x$
($v_{\mathrm{ref}}$: reference propagation velocity; $\Delta t$: sampling interval; $\Delta x$: inter-channel spacing).
This maps the full spatial aperture of the segment to an equivalent temporal window at the reference velocity.
Multi-scale RoIAlign is then applied again to the adjusted mask RoIs to produce mask-branch inputs.

\paragraph{Box head.}
Candidate regions are extracted from the feature pyramid using multi-scale RoIAlign with an output resolution of $7\times7$. 
The pooled features are processed by two fully connected layers and then fed into two prediction heads: a classifier that produces class logits (including background) and a box regressor that refines bounding-box coordinates. 
During inference, bounding boxes are decoded, filtered using a confidence threshold, and processed with class-wise NMS to produce the final detections.

\paragraph{Mask head.}
After temporal localization, multi-scale RoIAlign extracts a $42\times42$ feature region for each proposal.
The mask head consists of four $3\times3$ convolutional layers (256 channels, batch normalization, and ReLU activation), followed by a $2\times2$ transposed convolution for upsampling and a final $1\times1$ convolution that outputs per-class mask logits. 
During training, a weighted binary cross-entropy loss is used, with a Gaussian-smoothed weighting map that emphasizes pixels near signal arrivals and regions with higher confidence.

